\documentclass[pdflatex,sn-mathphys-num]{sn-jnl}

\usepackage{graphicx}%
\usepackage{multirow}%
\usepackage{amsmath,amssymb,amsfonts}%
\usepackage{amsthm}%
\usepackage{mathrsfs}%
\usepackage[title]{appendix}%
\usepackage{xcolor}%
\usepackage{textcomp}%
\usepackage{manyfoot}%
\usepackage{booktabs}%
\usepackage{algorithm}%
\usepackage{algorithmicx}%
\usepackage{algpseudocode}%
\usepackage{listings}%

\theoremstyle{thmstyleone}%

\theoremstyle{thmstyletwo}%

\theoremstyle{thmstylethree}%
\newtheorem{definition}{Definition}%

\begin{document}

%
\title[Article Title]{
Generalized L-Modularity for Community Detection Beyond Simple Temporal Networks
}

 \author[*]{\fnm{Victor} \sur{Brabant}}\email{victor.brabant@liris.cnrs.fr}

 \author{\fnm{Angela} \sur{Bonifati}}\email{angela.bonifati@univ-lyon1.fr}

 \author{\fnm{Rémy} \sur{Cazabet}}\email{remy.cazabet@univ-lyon1.fr}


\affil{Université Claude Bernard Lyon 1, CNRS, INSA Lyon, LIRIS, UMR5205, \orgaddress{\city{Villeurbanne}, \postcode{F-69622}, \country{France}}}



\abstract{
Detecting communities in networks is essential for understanding the mesoscopic organization of complex systems. Interactions in most real-world networks evolve over time and exhibit diverse modalities: instantaneous events, continuous contacts that persist over intervals, and delayed interactions where source and destination are temporally separated, as observed in transportation processes. Additionally, interactions may be directed, weighted, or involve multiple node types.
Existing methods for community detection in temporal networks typically handle only limited subsets of these features. When applied to real-world data, they often rely on simplifying transformations, such as aggregating interactions into time windows, projecting multipartite structures onto unipartite graphs, or ignoring edge directions and weights, leading to a loss of information.
In this work, we generalize Longitudinal Modularity (L-Modularity) and the LAGO algorithm into a unified framework for dynamic community detection in complex link streams.
Experiments on three real-world datasets demonstrate that our approach discovers meaningful communities in temporal networks with diverse interaction types.
}


\maketitle  

\section{Introduction}

Complex networks provide a fundamental framework for modeling a wide range of real-world systems, from social interactions and communication networks to transportation systems, financial markets, and biological processes~\cite{statistical_albert_2001,structure_newman_2003}.
A central problem in this domain is the detection of \emph{community structure}: groups of nodes that interact more densely among themselves than with the rest of the network~\cite{fortunato2016community}.
Communities reveal mesoscopic organization, reduce the complexity of large-scale networks, and are essential for understanding underlying dynamics.

For static networks, community detection is a mature and extensively studied problem.
While numerous approaches exist, one of the most common is Modularity~\cite{newman2004finding}, using efficient greedy algorithms such as Louvain~\cite{Blondel_2008} or Leiden~\cite{Traag_2019} to optimize it.

Real-world systems, however, are inherently dynamic.
Interactions evolve over time and take qualitatively different forms: an email is sent at a precise instant (\emph{instantaneous}), two colleagues share an office over a time interval (\emph{continuous}), and a bus departs from one stop and arrives at another after a measurable delay (\emph{delayed}).

The dominant paradigm for community detection in temporal networks extends static methods to sequences of graphs known as snapshots.
Among the most popular methods, multilayer modularity~\cite{Mucha_2010} jointly optimizes community structure across all snapshots.
While effective for slowly evolving networks~\cite{cazabet2020evaluating},
snapshot-based methods fundamentally require choosing a time scale for aggregation --- a choice that is often arbitrary and typically causes a loss of temporal precision~\cite{effects_krings_2012}.

To overcome this limitation, a more general approach works directly on event sequences, without any prior transformation.
The \emph{link stream} model~\cite{latapy2018stream} formalizes such data as a set of timestamped interactions, enabling community detection at the finest temporal resolution of the observed data.

Within this framework, \emph{Longitudinal Modularity} (L-Modularity)~\cite{Brabant2025-rw} generalizes classical modularity to link streams. It relies on the LAGO algorithm~\cite{brabant2025discoveringcommunitiescontinuoustimetemporal} to identify relevant temporal partitions, allowing nodes to join and leave communities at precise interaction times.

Existing approaches address only a limited subset of interaction features, typically focusing on \emph{simple} --- undirected and unweighted --- interactions that are either instantaneous or continuous, among homogeneous nodes.
However, most real-world systems do not fit these restrictive settings.
Financial transactions carry a sender, a receiver, and an amount (directed, weighted).
Online platforms connect heterogeneous entities --- users, posts, hashtags --- forming multipartite structures.
Furthermore, as illustrated above, temporal interactions may be instantaneous, continuous, or delayed.
To our knowledge, no existing method for community detection in temporal networks is able to deal with such data directly, without first performing a destructive transformation into a temporal network composed of simpler interactions.

In this paper, we generalize both L-Modularity and LAGO to fill this gap.
We first establish the formal setting by introducing a \emph{generalized link stream} model that accommodates weighted and directed interactions, multipartite node structures, and three interaction modalities: instantaneous, continuous, and delayed. Our two main contributions are then:
\begin{enumerate}
    \item \textbf{Generalized Longitudinal Modularity.}
    We extend L-Modularity to evaluate dynamic community structures in generalized link streams, incorporating weighted, directed, multipartite, continuous, and delayed interactions into a single quality function that reduces to the original formulation in the case of simple link streams.

    \item \textbf{Generalized LAGO.}
    We adapt the LAGO optimization procedure to maximize the generalized L-Modularity, introducing \emph{active time-segment nodes} for continuous interactions and updating neighborhood and local-move definitions to account for the full range of interaction types.
\end{enumerate}

We validate the framework on three real-world datasets spanning the interaction types considered: a directed and weighted voting network (Eurovision Song Contest), a transportation network with delayed and continuous interactions (Citi Bike), and a bipartite social network (Bluesky).
In each case, LAGO uncovers dynamic communities consistent with known domain knowledge, and we show that modeling choices --- direction, weights, interaction modality --- meaningfully affect the resulting community structure.

The remainder of this paper is organized as follows.
Section~\ref{sec:relwork} reviews related work.
Section~\ref{sec:linkstream} introduces the generalized link stream model.
Section~\ref{sec:longmod} presents the generalized Longitudinal Modularity.
Section~\ref{sec:lago} describes the adapted LAGO optimization procedure.
Section~\ref{sec:applications} illustrates the approach on real-world datasets.
Section~\ref{sec:discussion} concludes and outlines future directions.

\section{Related work}\label{sec:relwork}

We first identify the gap in the literature in terms of methods to detect temporal communities in non-simple temporal graph settings, then review how this problem has been solved for static networks by various authors who proposed extensions of both the static modularity and Louvain method.

\subsection{Community Detection in Temporal Networks}

This section reviews the main approaches for community detection in temporal networks, with a focus on their ---partial--- ability to handle non-simple interactions such as weighted, directed, multipartite, instantaneous, continuous, and delayed interactions.

\subsubsection{Snapshot Sequences}

A widely adopted paradigm represents a dynamic network as an ordered sequence of static snapshots. Static community detection methods can be applied independently to each time window, sometimes adding an implicit constraint to enforce temporal consistency, such as biasing the network at time $t$ by the community structure found at $t-1$~\cite{cazabet2020evaluating}.  
While these approaches are usually tested with standard static community detection algorithms such as Louvain, some of them are method-agnostic: since they independently apply static algorithms on each snapshot, they can support non-simple network modalities such as multipartite networks or directed edges. 
Although this has been little explored in the literature, we can find at least one instance of this approach to detect bipartite temporal partitions~\cite{tempweightbip}. The authors used as a static partition objective the bipartite modularity, limiting instability by using the Implicit-Global approach~\cite{cazabet2020evaluating} on the Louvain algorithm.

However, these methods using independent detections on each snapshot suffer from the smoothness problem~\cite{rossetti2018community}, leading to unstable temporal partitions. 

Various approaches have been proposed to solve this instability problem by explicitly enforcing both high topological quality and temporal stability. 
One of the best known has been introduced by Mucha et al.~\cite{Mucha_2010}: it incorporates inter-layer coupling edges between snapshots, and then uses an adapted Modularity to uncover communities containing nodes in multiple snapshots in a single optimization process. A parameter $\omega$ controls the strength of temporal coupling and thus the persistence of communities across layers.
As an alternative, multilayer Infomap~\cite{infomap_multilayer} uses the same mechanism with its own flow-based objective, relying on random walks evolving across layers. 
These approaches can natively accommodate weighted and directed networks.

Stochastic block models (SBMs)~\cite{HOLLAND1983109} provide a complementary framework for dynamic networks, where communities are found through the inference of the parameters of a generative statistical model, usually maximizing likelihood. In dynamic SBMs, node memberships typically evolve over time according to parametric mechanisms, most often Markov processes~\cite{detecting_yang_2011}, or are inferred via state-space formulations with time-varying connectivity parameters~\cite{dynamic_xu_2014}. Extensions based on hidden Markov models further allow probabilistic transitions between groups in weighted and directed networks~\cite{statistical_matias_2017}.
However, snapshot-based SBMs remain limited: they usually assume a fixed number of groups and rely on restrictive parametric dynamics. In addition, they do not naturally account for interactions with duration or delay, and require adaptation for multipartite networks, which limits their applicability to many real-world temporal networks.

To the best of our knowledge, no snapshot-based approach handles delayed or continuous interactions. In principle, coupling-graph methods could represent delayed interactions via cross-snapshot edges, but this has not been explored.

Overall, these methods suffer from the well-known limitations of snapshot-based approaches~\cite{latapy2018stream,rossetti2018community,Brabant2024-lb,cazabet2020evaluating}. They are restricted to progressively evolving networks, where a well-defined graph exists at each snapshot. For rapidly evolving networks, they instead require a destructive transformation of interaction flow into snapshots~\cite{effects_krings_2012}.

A complementary class of methods concerns slowly evolving temporal networks with no need for a snapshot representation, i.e., graphs with interactions defined over intervals typically having long durations. The most popular such methods are iLCD~\cite{cazabet2010detection}, TILES~\cite{rossetti2017tiles}, and OLCPM~\cite{boudebza2018olcpm}. However, these methods are not able to deal with fast-evolving networks, and are based on local heuristics, i.e., without a well-defined objective function to optimize. They also cannot work with weighted, directed or bipartite temporal networks.

\subsubsection{Link Streams and Stream graphs}

An alternative line of work models temporal networks as sequences of interaction events, thereby preserving exact temporal resolution. The \emph{link stream} and \emph{stream graph} formalisms~\cite{latapy2018stream} represent the system as a set of timestamped interactions and naturally encompass both rapidly evolving dynamics and more progressively evolving networks.

A few methods have been proposed in this setting, but most address only a subset of the requirements for fully dynamic community detection. 

Continuous-time extensions of stochastic block models (SBMs)~\cite{exact_corneli_2016, block_corneli_2016, matias2018semiparametric, modelling_peixoto_2017} partition nodes into latent groups and model time-varying interaction rates, often via change-point processes. In most cases, node memberships are static; an exception is~\cite{modelling_peixoto_2017}, where memberships evolve according to a Markov process. These models generally assume unipartite, undirected, and unweighted interactions, and do not capture continuous or delayed edges;~\cite{modelling_peixoto_2017} partially relaxes these assumptions by allowing directed interactions.

LSCPM~\cite{baudin_et_al:LIPIcs.TIME.2023.3} relies on a local heuristic requiring continuous interactions, which necessitates artificially extending instantaneous events, therefore introducing an arbitrary transformation of the data. It does not support weighted, directed, or bipartite networks.

The flow stability framework~\cite{bovet2022flow} provides a dynamics-based method that derives communities from a diffusion process evolving on the temporal network. Unlike snapshot methods, it respects the temporal ordering of edges
. By varying the speed of the diffusion process, it can also reveal community organization at different dynamical scales. However, community transitions are restricted to a finite set of predefined time points chosen by the user, and the method assumes instantaneous or continuous interaction dynamics, which limits its applicability to networks with delayed interactions. Furthermore, it does not address multipartite structures. 

Brabant et al.~\cite{Brabant2025-rw} introduced \emph{Longitudinal Modularity} (L-Modularity), a quality function designed for link streams. It evaluates dynamic community assignments by comparing observed intra-community interactions against a longitudinal null model.  
The LAGO algorithm~\cite{brabant2025discoveringcommunitiescontinuoustimetemporal} optimises L-Modularity using a Louvain-like heuristic adapted to link streams, allowing nodes to join or leave communities at exact interaction times. This approach is limited to simple instantaneous interactions.

\subsection{Modularity and the Louvain Method in Static Networks}

This section reviews the main extensions and limitations of the modularity quality function~\cite{newman2004finding} and of the Louvain optimization method~\cite{Blondel_2008}, including their adaptations to weighted, directed, and bipartite static networks.

In this work, we focus on the modularity--Louvain framework, as it provides a clear separation between the definition of a quality function and its optimization. This separation allows each component to be extended independently to incorporate additional network features, in particular along temporal dynamics.

\subsubsection{Modularity and its Extensions}

Modularity~\cite{newman2004finding} is one of the most widely used quality functions for analyzing community structure in networks. Despite well-documented limitations, such as the resolution limit, it remains popular due to its intuitive interpretation, simple formulation, good results in practice, and the availability of efficient optimization heuristics. Here, we first introduce its original definition, and then several extensions to non-simple graphs.

Consider a network with node set $V$ and adjacency matrix $A$, where $A_{uv}$
denotes the weight of the edge from $u$ to $v$ (or simply $1$ if the network is unweighted).
The degree of a node is $k_u = \sum_v A_{uv}$, and the total edge weight is
$m = \frac{1}{2}\sum_u k_u$.
A community structure is a partition $\mathcal{C}$ of~$V$ into groups with a high concentration of intra-group edges relative to inter-group edges.

Modularity evaluates the quality of a given partition by comparing the observed edge density within communities to its expected value under a degree-preserving null model (the configuration model).
Introducing a resolution parameter $\gamma$~\cite{PhysRevE.74.016110}, which controls the characteristic size of detected communities and partially alleviates the resolution limit~\cite{Fortunato_2007}, modularity is defined as
\begin{align}\label{eq:mod}
    Q = \frac{1}{2m} \sum_{C \in \mathcal{C}} \sum_{u,v \in C}
    \left[ A_{uv} - \gamma\frac{k_u k_v}{2m} \right].
\end{align}
High values of $Q$ indicate a statistically surprising concentration of edges within communities. When $A_{uv}$ encodes interaction strengths rather than binary indicators, the same formulation naturally extends to \emph{weighted networks}.

For \emph{directed networks}, the adjacency matrix is asymmetric ($A_{uv} \neq A_{vu}$ in general) and each node has an out-degree $k_u^{\mathrm{out}} = \sum_v A_{uv}$ and an in-degree $k_u^{\mathrm{in}} = \sum_v A_{vu}$, with total weight $w = \sum_u k_u^{\mathrm{out}} = \sum_u k_u^{\mathrm{in}}$.
Leicht and Newman~\cite{Leicht_2008} proposed a tailored null model based on separate in- and out-degree sequences:
\begin{align}\label{eq:mod:dir}
    Q = \frac{1}{w} \sum_{C \in \mathcal{C}} \sum_{u,v \in C}
    \left[ A_{uv} - \gamma\frac{k_u^{\mathrm{out}} k_v^{\mathrm{in}}}{w} \right].
\end{align}
The undirected formulation is recovered by treating each undirected edge as two reciprocal directed edges, so that $k_u^{\mathrm{in}} = k_u^{\mathrm{out}} = k_u$.

Some networks involve nodes of different kinds --- such as users and items, or genes and diseases --- and interactions only occur between different kinds.
For such \emph{multipartite networks}, the node set is partitioned into disjoint types
$\{V_1, \dots, V_p\}$ and edges exist only between nodes of different types
($A_{uv} = 0$ whenever $u, v \in V_i$); the \emph{bipartite} case corresponds to $p = 2$.
Barber~\cite{Barber_2007} introduced a modularity formulation that restricts the null model to respect this constraint: since no edges can exist within the same type, the expected number of intra-type edges is set to zero, and the null model only redistributes edges across types. This is achieved by setting $b_{uv} = 1$ when $u$ and $v$ belong to different types and $0$ otherwise, effectively masking same-type pairs from the expectation term. The weighted, directed, and multipartite extensions then combine into a single unified formulation:
\begin{align}\label{eq:mod:unified}
    Q = \frac{1}{w} \sum_{C \in \mathcal{C}} \sum_{u,v \in C}
    \left[ A_{uv} - \gamma\, b_{uv} \frac{k_u^{\mathrm{out}} k_v^{\mathrm{in}}}{w} \right].
\end{align}
Setting $b_{uv} \equiv 1$, $k_u^{\mathrm{in}} = k_u^{\mathrm{out}} = k_u$ and therefore $w = 2m$ recovers \eqref{eq:mod}. This formulation is the static foundation from which the generalized Longitudinal Modularity introduced in Section~\ref{sec:longmod} is derived.

Other extensions of modularity exist, notably for signed networks, overlapping communities, and higher-order structures. We believe that incorporating them into the temporal setting would require dedicated work and is left for future research.

\subsubsection{The Louvain Method and Extensions}\label{sec:rel:stat:louv}

Modularity defines what a good community structure is, but finding one requires an optimization algorithm. Since maximizing modularity is NP-hard, practical methods rely on greedy heuristics.

The Louvain method~\cite{Blondel_2008} is the standard algorithm for maximizing modularity at scale. It alternates between a local node-reassignment phase, in which each node greedily joins the community of a neighbor that yields the highest gain in \eqref{eq:mod}, and a community-aggregation phase that contracts each community into a single node.
A later improvement, the Leiden algorithm~\cite{Traag_2019}, adds a refinement phase that guarantees well-connected communities and typically yields higher-quality results.
Both algorithms accommodate weighted and directed networks without structural modification, since the network type only affects the gain formula used in the reassignment phase~\cite{directed_dugu_2015}.
For bipartite networks, the bi-Louvain variant~\cite{novel_zhou_2018} performs type-aware aggregation to prevent merging nodes of different types, and the extension to general multipartite structures follows the same principle.

\section{Generalizing Link Streams Beyond Simple Interactions}\label{sec:linkstream}

\textit{Link streams} model temporal interactions as series of triplets $(u,v,t)$, each representing an \textit{interaction} between nodes $u$ and $v$ at time $t$. These interactions are typically assumed to be instantaneous. In contrast to representations based on time-ordered sequences of static graphs, link streams can represent flows of interactions without the need to aggregate them over a time window, or having to choose a collection frequency.

In this section, we introduce the concept and definition of \emph{generalized link stream}, mainly inspired by \cite{latapy2018stream,Latapy2019}, which will be used throughout the article. Although this definition does not cover all possible variants of link streams, we adopt the term \emph{generalized link stream} in contrast to what we call \emph{simple link streams}, on which the original L-Modularity is defined.

Real-world temporal data often involves richer interaction mechanisms, including \textbf{weights}, \textbf{directions}, or \textbf{multipartite} node structures. For example, financial transactions are often modeled as having a \emph{sender} (source), sending some \emph{amount} to a \emph{receiver} (destination) at a particular time; online social networks may record events as having a \emph{user} associated with a \emph{hashtag} at a specific time. Weights, directionality and heterogeneity of nodes have already been widely studied in the literature on static networks; we will extend these works to link streams.

In generalized link streams, we will also introduce additional interaction modalities that are specific to the temporal setting. Indeed, interactions may be instantaneous (e.g., sending an email), may persist over a time interval (e.g.,\ phone calls), or may involve delays between departure and arrival (e.g., transportation systems).
An \textbf{instantaneous interaction} occurs at a single time $t$ and corresponds to the classical link stream event $(u,v,t)$.
A \textbf{continuous interaction} persists over a time interval $[t_s, t_e)$ and naturally represents sustained phenomena such as co-location or ongoing communication. Such an interaction is denoted $(u, v, t_s, t_e)$.
Many systems involve \textbf{delayed interactions}, where an interaction departs from node $u$ at time $t_s$ and reaches node $v$ at a later time $t_d$, for example airplane flights or shipments in supply chains. These interactions are denoted $(u, v, t_s ,t_d)$.

Instantaneous interactions correspond to the special case $t_s = t_d$. We refer to both instantaneous and delayed interactions collectively as \textbf{timestamp-based interactions} (since each is fully characterized by one or two discrete timestamps), as opposed to \textbf{interval-based interactions} corresponding to continuous interactions. We adopt this terminology to avoid ambiguity with the continuous or discrete nature of the time domain of the network. 

\subsection{Definition of a Generalized Link Stream}

In this section, we define the generalized link stream, of which the simple link stream is a particular case.

\begin{definition}\label{def:linkstream:complex}
A \textbf{link stream} is defined as
\[
L = \left( T, V, E \right),
\]
where $T \subset \mathbb{R}$ (resp.\ $T \subset \mathbb{Z}$) is a continuous (resp.\ discrete) time interval, $V$ is a finite set of nodes, and $E$ is a finite set of temporal interactions.

Each interaction is \textbf{directed}, with the ordered pair $(u,v)$ encoding the direction from node $u$ to node $v$, and carries a \textbf{weight} $w \in \mathbb{R}_+$, assumed to be $1$ when not specified. Undirected interactions are treated as two reciprocal directed interactions.

Depending on the temporal nature of the interactions, the edge set $E$ takes one of two forms.

\paragraph{Timestamp-based interactions}
\begin{align}
E = \{(u, v, t_s, t_d, w) \in V^2 \times T^2 \times \mathbb{R}_+ \;:\; t_s \le t_d\},
\end{align}
where $t_s$ and $t_d$ denote the departure time from node $u$ and the arrival time at node $v$, respectively. Instantaneous interactions correspond to the special case $t_s = t_d$. We refer to both instantaneous and delayed interactions collectively as \textbf{timestamp-based interactions}.

\paragraph{Interval-based interactions}
\begin{align}
E = \{(u, v, t_s, t_e, w) \in V^2 \times T^2 \times \mathbb{R}_+ \;:\; t_s \le t_e\},
\end{align}
where $[t_s, t_e)$ is the time interval during which the interaction persists. This corresponds to what was earlier referred to as continuous interactions.

We adopt the terms timestamp-based and interval-based to avoid ambiguity with the continuous or discrete nature of the time domain~$T$. A given link stream is assumed to contain exclusively one type or the other.

\paragraph{Multipartite link streams}

When the network is \textbf{multipartite}, the node set is written
\[
V = \bigcup_{i=1}^{P} V_i,
\]
where $\{V_i\}_i^P$ is a partition of $V$. Interactions are only allowed between different parts, i.e.\ $u, v \in V_i \;\Rightarrow\; (u,v, \cdot, \cdot, \cdot) \notin E$. When $P = 2$, the link stream is said to be bipartite.
\end{definition}

\subsection{Generalizing Useful Definitions on Link Streams}
To redefine the L-Modularity over a generalized link stream, we need to introduce some notations common to the different types of generalizations.

\subsubsection{Interactions Over a Period}


\paragraph{Timestamp-based interactions}
We define
\[
I_{uv, T'T''} = \{ (u,v, t_s,t_d, w) \in E \text{ such that } (t_s,t_d) \in T' \times T''\}
\]
the set of interactions leaving $u$ during the period $T'$ and arriving at $v$ during the period $T''$.
The \textbf{number} of interactions $L$, and the \textbf{total weight} of interactions $W$ are defined respectively as
\[
L_{uv, T'T''} = |I_{uv, T'T''}|, \quad W_{uv, T'T''} = \sum_{\substack{(u,v, t_s, t_d, w) \\ \in \; I_{uv, T'T''}}} w.
\]
For \textbf{instantaneous interactions}, we simply write
\[
I_{uv, T'} = I_{uv, T'T'}, \quad L_{uv, T'} = L_{uv, T'T'}, \quad \text{and }  W_{uv, T'} = W_{uv, T'T'}.
\]

\paragraph{Interval-based interactions}
We define the instantaneous interaction intensity
\[
uv_t = \sum_{\substack{(u,v,t_s,t_e,w)\in E \\ t_s \le t < t_e}} w,
\]
which represents the total weight of interactions from $u$ to $v$ at time $t$.
The \textbf{total weight} of interactions from $u$ to $v$ over an interval $T' \subseteq T$ is
\[
W_{uv,T'} =
\int_{t \in T'} uv_t\,dt
\quad \text{if } T \subset \mathbb{R},
\qquad
W_{uv,T'} =
\sum_{t \in T'} uv_t
\quad \text{if } T \subset \mathbb{Z}.
\]

\begin{figure}[htbp]
    \centering
    \begin{minipage}[b]{0.49\linewidth}
        \centering
        \includegraphics[width=\linewidth]{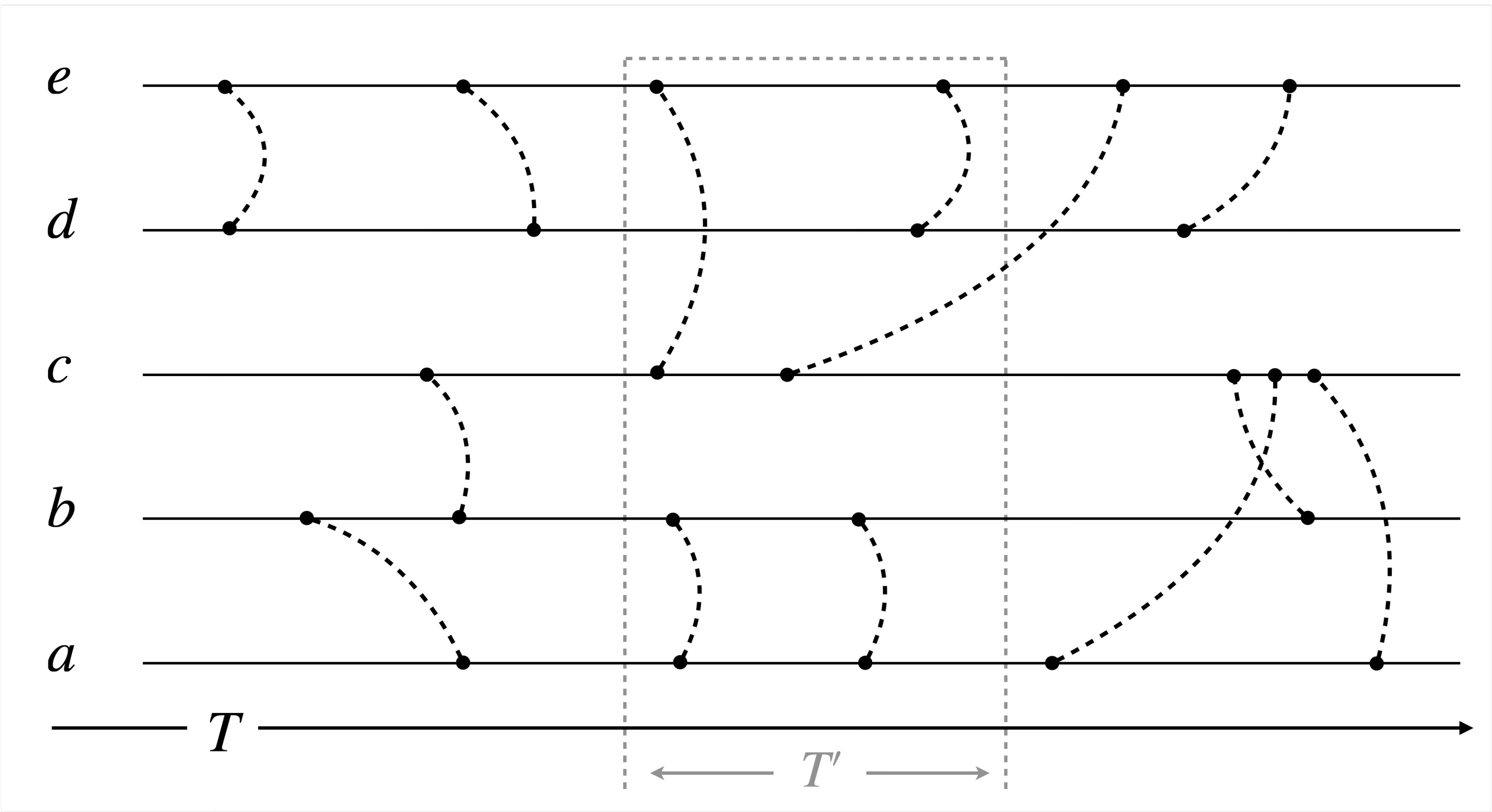}
        \vspace{2pt}
        \small (a) Timestamp-based interactions
    \end{minipage}
    \hfill
    \begin{minipage}[b]{0.49\linewidth}
        \centering
        \includegraphics[width=\linewidth]{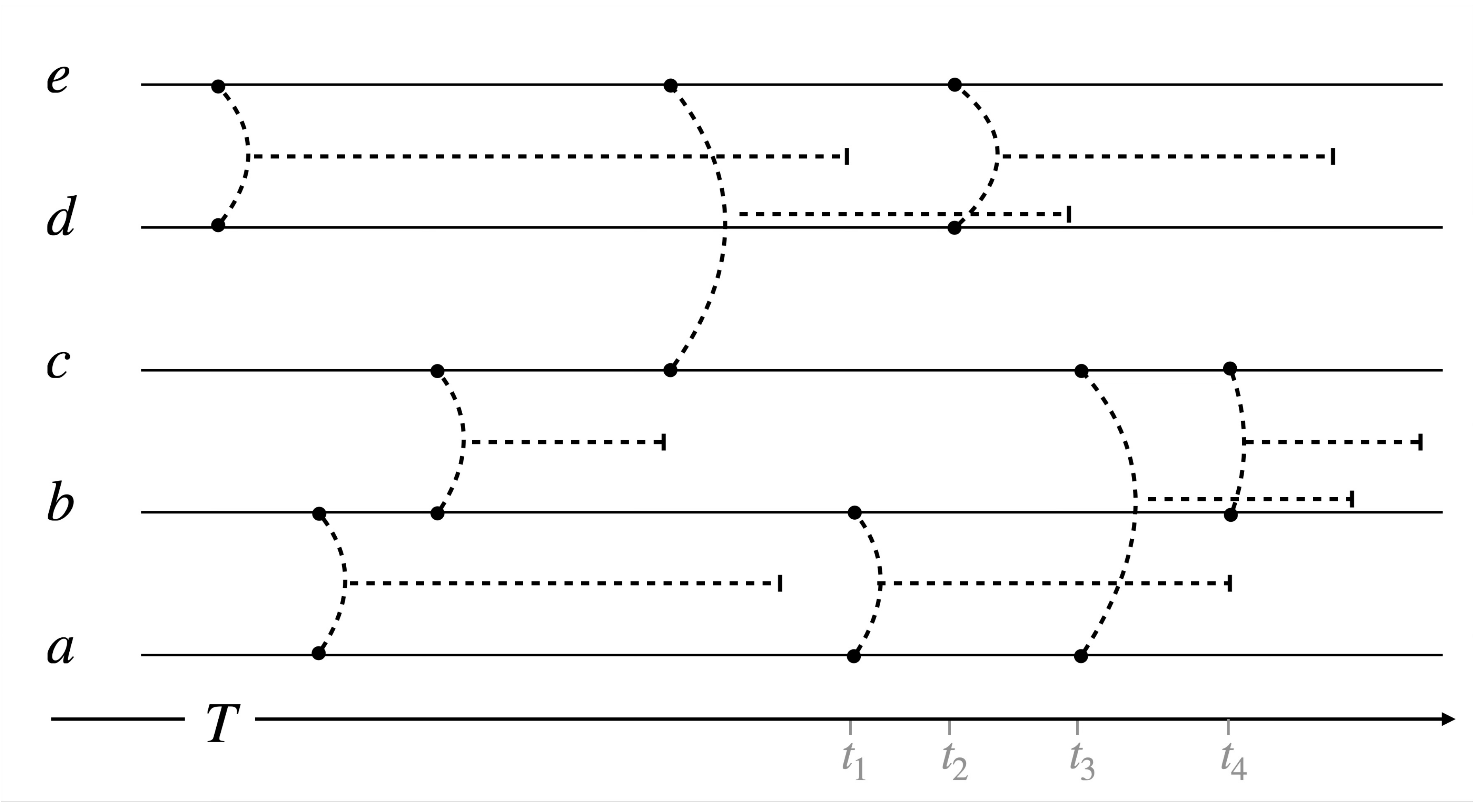}
        \vspace{2pt}
        \small (b) Interval-based interactions
    \end{minipage}

    \vspace{4pt} 
    \caption{\textbf{Link stream representations.} 
    Interactions are unweighted and undirected over the node set $V=\{a,b,c,d,e\}$. Nodes are shown on the vertical axis and time $T$ on the horizontal axis. 
    \textbf{Panel (a):} 13 interactions, including both instantaneous and delayed events, e.g., $W_{ce,T'} = 1$ and $W_{ce} = 2$. 
    \textbf{Panel (b):} 8 interactions; durations are represented as horizontal segments; for example, $ab_{t_2} = 1$, $bd_{t_1} = 0$, and $W_{ac,[t_1,t_4)} = t_4 - t_3$.}
    \label{fig:ls:examples}
\end{figure}

\paragraph{General setting}
In both timestamp-based and interval-based settings we denote the \textbf{total interaction weight} from $u$ to $v$ over the whole observation period as
\[
W_{uv} = W_{uv, T}.
\]
Figure~\ref{fig:ls:examples} illustrates both timestamp-based and interval-based link streams with examples of the notations introduced in this section.

\subsubsection{Total In-Out Degrees}
To define the null model of L-Modularity, we need to use the total degree of nodes over the whole period of the link stream. 
Similarly to directed weighted graphs in the static setting, the in- and out-degrees are defined as
\[
k_u^{\mathrm{in}} = \sum_{v \in V} W_{vu},
\qquad
k_u^{\mathrm{out}} = \sum_{v \in V} W_{uv}.
\]
Finally, the total weight of the link stream is
\[
w = \sum_{u\in V} k_u^{\mathrm{in}}
  = \sum_{u\in V} k_u^{\mathrm{out}}
  = \sum_{u v\in V^2} W_{uv},
\]
and the number of interactions is $m = |E|$. 

We consider that interactions in undirected link streams should be equivalent to having directed interactions in both directions. As a consequence, we have
\[
k_u = k_u^{\mathrm{in}}  = k_u^{\mathrm{out}}, \quad w = \sum_{u\in V} k_u,
\]
respectively the undirected degree of node $u$ and the weight of the network. In the unweighted and undirected case, $w = 2|E| = 2m$.

These notations allow to express all concepts used in the following sections, regardless of whether $T$ is continuous or discrete, and independently of the interaction temporal type (timestamp-based or interval-based) or additional features introduced above. 

\section{L-Modularity Beyond Simple Link Streams}\label{sec:longmod}

Brabant et al.~\cite{Brabant2025-rw} introduced the \emph{Longitudinal Modularity} (L-Modularity) to evaluate dynamic community structures in simple instantaneous link streams, allowing exact temporal precision for nodes entering and leaving communities.  
In this section, we extend L-Modularity beyond simple instantaneous link streams (Sec.~\ref{sec:lmod:extention}). We first specify the notion of dynamic communities considered in this work (Sec.~\ref{sec:lmod:dyncom}), and then recall the original definition of L-Modularity (Sec.~\ref{sec:lmod:original}).

\subsection{Dynamic Communities}\label{sec:lmod:dyncom}

In static networks, communities are typically defined as a partition of the node set, i.e., a collection of sets such that each node belongs to exactly one community. In this article, we follow the definition of dynamic communities introduced in \cite{Brabant2025-rw}, which preserves the principle of non-overlapping communities while allowing two key temporal behaviors: (i) community membership may evolve over time, and (ii) nodes may remain outside of any community during periods of inactivity. 
\begin{definition}
Let $L = (T, V, E)$ be a general link stream. A \textbf{dynamic community structure} over $L$ is defined as a collection of \textbf{non-empty} and \textbf{mutually exclusive} communities composed of sets of node--time-segment pairs 
\[ 
\left\{\left(u_1, [t_1, t_1')\right), \left(u_2, [t_2, t_2')\right), \ldots, \left(u_2, [t_2'', t_2''')\right), \left(u_3, [t_3, t_3')\right), \ldots\right\}
\] 
such that, for any node $u$, the time segments associated with $u$ do not overlap. This definition is compatible with both discrete and continuous time domains $T$ and any type of interactions $E$.
\end{definition}
We introduce the following notations. For a node $u$ and a community $C$,
\[
T_{u \in C} = \bigcup_{(u,[t_i,t_{i+1})) \in C} [t_i,t_{i+1})
\]
denotes the set of times during which node $u$ belongs to community $C$.
\[
L_{uv \in C} = L_{uv, T_{u \in C}T_{v \in C}}, \quad W_{uv \in C} = W_{uv, T_{u \in C}T_{v \in C}}
\]
denote respectively the number of interactions (for timestamp-based link streams only) and the cumulative weight of interactions from node $u$ to node $v$ occurring while both nodes belong to community $C$.

\[
T_{C} = \bigcup_{u \in V} T_{u \in C}
\]
denotes the existence time of community $C$.

\subsection{L-Modularity for Simple Temporal Networks}\label{sec:lmod:original}

To evaluate the quality of dynamic communities in simple instantaneous link streams, Brabant et al.~\cite{Brabant2025-rw} introduced the \emph{Longitudinal Modularity} (L-Modularity), an extension of the classical modularity measure. Similar to its static counterpart~\eqref{eq:mod}, L-Modularity compares the fraction of observed interactions within communities with the expected fraction under a \emph{longitudinal random null model}. The larger this difference, the higher the L-Modularity.

The expected number of interactions $\mathbb{E}_{\star}\left[L_{uv\in C}  \right]$ between nodes $u$ and $v$ within a dynamic community $C$ is obtained by weighting the configuration null model of the static setting by a temporal factor:
\begin{align}
    \mathbb{E}_{\star}\left[L_{uv \in C} \right] = \dfrac{k_{u} k_{v}}{2m} \mathbb{T}_{uv, C}^{\star},
\end{align}
where $\mathbb{T}_{uv, C}^{\star}$ correspond to two variants of longitudinal random null model: the \textbf{Joint-Membership (JM)} and the \textbf{Mean-Membership (MM)}.

JM \eqref{eq:le:jm} expects the overall structure of the community to be stationary or nearly stationary, meaning that most nodes remain in the same community throughout their lifetime: 
\begin{align}\label{eq:le:jm}
    \mathbb{T}_{uv, C}^{\mathrm{JM}} = \dfrac{|T_C|}{|T|} \mathbf{1}_{|T_{u\in C}||T_{v\in C}| > 0}.
\end{align}
In contrast, the MM formulation \eqref{eq:le:mm} is more permissive and allows more frequent changes in node affiliations during the lifetime of a community:
\begin{align} \label{eq:le:mm}
    \mathbb{T}_{uv, C}^{\mathrm{MM}} = \frac{\sqrt{|T_{u \in C}| |T_{v \in C}|}}{|T|}.
\end{align}

L-Modularity also includes a temporal regularization term that quantifies the temporal smoothness of dynamic community structure.
This term is based on the sum of the \textbf{Community Switch Count} (CSC) over all nodes, $\sum \eta_u$, where $\eta_{u}$ denotes the CSC of node $u$. The CSC measures how frequently a node changes its community membership over time. Specifically, $\eta_u$ is the number of communities visited by node $u$ minus one, and zero if the node never leaves a community. 

The regularization term is weighted by a parameter $\omega \geq 0$, which controls the temporal smoothness. When $\omega=0$, L-Modularity favors communities with instantaneous lifetimes. Larger values of $\omega$ encourage temporal stability in node memberships.

Finally, the L-Modularity formulation of a dynamic community set $\mathcal{C}$ on a simple link stream $L= (T, V, E)$ is
\begin{align}
         Q_{\star } (L, \mathcal{C}, \omega) = 
         & \frac{1}{2m} \sum_{C \in \mathcal{C}} \sum_{u,v \in V^2} \left[ L_{uv \in C} - \dfrac{k_{u} k_{v}}{2m} \mathbb{T}_{uv, C}^{\star}\right] \nonumber \\
         & - \frac{\omega}{2m} \sum_{u \in V} \eta_{u}(\mathcal{C}),
\end{align}
where $\star =$ JM or MM, and $\omega \geq 0$ a time smoothness parameter. 

\subsection{L-Modularity Beyond Simple Temporal Networks}\label{sec:lmod:extention}

We extend L-Modularity to evaluate dynamic community structures in temporal networks that may be weighted, directed, continuous, delayed, or defined over multipartite node sets. To our knowledge, this constitutes the first quality function designed to evaluate dynamic community structures in such general temporal network settings. 
The section first discusses the main components of L-Modularity affected by this generalization, then introduces the formal definition, and finally discusses the role of meta-parameters as well as the choice of null model with respect to the different types of interactions.

\subsubsection{Conceptual Components}

The function relies on three key components:  
(i) the observed interactions within communities;  
(ii) the expected amount of interactions under a null model; and  
(iii) a temporal regularization term ensuring temporal smoothness of the community structure.

\paragraph{Observed interactions}

As defined in the last section, the observed weight of interactions from node $u$ to node $v$ within community $C$ is given by $W_{uv \in C}$. The term encapsulates weighted and directed interactions, is independent of whether the node set is unipartite or multipartite, and is compatible with both timestamp-based and interval-based interactions.

\paragraph{Expected interactions}

We generalize the expectation term by extending the configuration null model used in static modularity~\cite{newman2004finding} to settings that allow weighted, directed, and multipartite networks. Such generalizations are well established in the static case and rely on replacing node degrees with appropriate in/out strengths~\cite{Leicht_2008} and restricting interactions according to node types~\cite{Barber_2007}. We adopt these extensions while preserving the temporal component of the longitudinal null model.
The expected weight of interactions from node $u$ to node $v$ within community $C$ is defined as
\begin{align}
\mathbb{E}_{\star}\left[W_{uv \in C}\right]
= 
b_{uv}
\frac{k_u^{\mathrm{out}}k_v^{\mathrm{in}}}{w}
\,\mathbb{T}_{uv,C}^{\star},
\end{align}
where $\star=$ JM or MM and $b_{uv}=1$ if interactions between nodes $u$ and $v$ are allowed (i.e., compatible types in multipartite networks) and $0$ otherwise. In unipartite networks, $b_{uv} \equiv 1$.   
In the quality function (Definition~\ref{def:lmodularity:general}), this expectation is scaled by a resolution parameter $\gamma > 0$, following the static setting~\cite{PhysRevE.74.016110}, that controls the granularity of detected communities.

In this formulation, certain link stream features are explicit --- such as multipartite constraints and interaction directionality --- while others, including interaction weights and whether interactions are interval-based or timestamp-based, are incorporated implicitly through the aggregated quantities.

For multipartite structures, no interactions are expected between nodes of the same kind and the weight of interactions expected between nodes of different kinds is equivalent to the unipartite case. 

For directed networks, the expected weight of interactions from $u$ to $v$ is proportional to the out-strength of $u$ ($k_u^{\mathrm{out}}$) and the in-strength of $v$ ($k_v^{\mathrm{in}}$).

Note that the expected weight of interaction does not depend on the temporal distribution of the interactions themselves. The null model only incorporates temporal information through the longitudinal factor $\mathbb{T}_{uv,C}^{\star}$, which reflects the co-membership of nodes in a community over time.

Consequently, two interaction patterns with the same total interaction weight over a given time interval have the same expectation under the null model. 
For instance, two continuous interactions of duration $d$ are equally expected as a single interaction of duration $2d$. Similarly, five instantaneous interactions of weight $1$ have the same expectation as one interaction of weight $5$.
Temporal concentration of interactions therefore appears as statistically surprising relative to the null model, potentially revealing communities.

The framework is adaptable to alternative null models tailored to specific datasets or objectives.

\paragraph{Temporal smoothness}

The temporal regularization term penalizes frequent community switches. In the original formulation, for simple instantaneous interactions, it is normalized by $2m$, which approximates the number of time nodes involved in interactions: each interaction contributes two endpoints. This normalization prevents degenerate configurations where each interacting time node forms a separate community, which would drive the expectation term to zero and artificially maximize L-Modularity. 

A natural question is whether this normalization should depend on the nature of interactions in the temporal network.

We adopt the principle that the counting of community switches should remain independent of the structural complexity of the link stream. In other words, the regularization term should reflect only the frequency of switches, and not be biased by how interactions are encoded.

Directed, delayed, and multipartite temporal networks do not require a modification of the temporal regularization term, as they do not alter the definition of instantaneous time nodes or community switches; consequently, the normalization by $2m$ remains unchanged.

In the weighted setting, one may wish to penalize community switches proportionally to the weights of the interactions involved. For instance, switches associated with low-weight interactions could be considered less significant, whereas switches affecting high-weight interactions would have a stronger impact on the desired temporal smoothness of communities. However, incorporating such refinements would require a substantial reformulation of the regularization term, which we leave for future work. 

In the interval-based setting, although interactions are not represented as timestamped events, the regularization should still be normalized by an approximation of the number of time nodes are involved in interactions. In this context, $2m$ (i.e., twice the number of interactions) remains a convenient proxy. This choice ensures that the regularization term is comparable across different temporal representations while preserving its role in preventing degenerate solutions.

\subsubsection{Formal Definition}

\begin{definition}\label{def:lmodularity:general}
    Let $L = (T, V, E)$ be a general link stream. With the previous notations, the \textbf{general Longitudinal Modularity} of a dynamic community set $\mathcal{C}$ regarding a resolution parameter $\gamma > 0$ and time parameter $\omega \geq 0$ is given by:
    \begin{align}
         Q_{\star } (L, \mathcal{C}, \gamma, \omega) = 
         & \frac{1}{w} \sum_{C \in \mathcal{C}} \sum_{u,v \in V^2} \left[ W_{uv \in C} - \gamma b_{uv}\dfrac{k_u^{\mathrm{out}} k_v^{\mathrm{in}}}{w} \mathbb{T}_{uv, C}^{\star}\right] \nonumber \\
         & - \frac{\omega}{2m} \sum_{u \in V} \eta_{u}(\mathcal{C}),
    \end{align}
    where $\star = \mathrm{JM} \;or \;\mathrm{MM} $. $b_{uv}=1$ if nodes $u$ and $v$ belong to different types and $b_{uv}=0$ otherwise. When unipartite, $b_{uv} \equiv 1$. When the network is unweighted and undirected, $w=2m$, and $W_{uv\in C}$ reduces to the number of interactions $L_{uv\in C}$ if interactions are discrete. For undirected networks, each interaction can be treated as two reciprocal directed interactions, so that $k_u=k_u^{\mathrm{in}}=k_u^{\mathrm{out}}$.
    
    \label{generalizedLmodularity}
\end{definition}

\subsubsection{Meta-Parameters Influence}\label{sec:metaparam}

The resolution parameter $\gamma$ controls the expected density of interactions within communities. Lower values favor larger, less dense communities, while higher values favor smaller, denser ones. The temporal penalty $\omega$ regulates community stability over time: lower values allow frequent community changes, while higher values encourage temporally stable partitions.  

\begin{figure}[htbp]
    \centering
    \begin{minipage}[b]{0.48\linewidth}
        \centering
        \includegraphics[width=\linewidth]{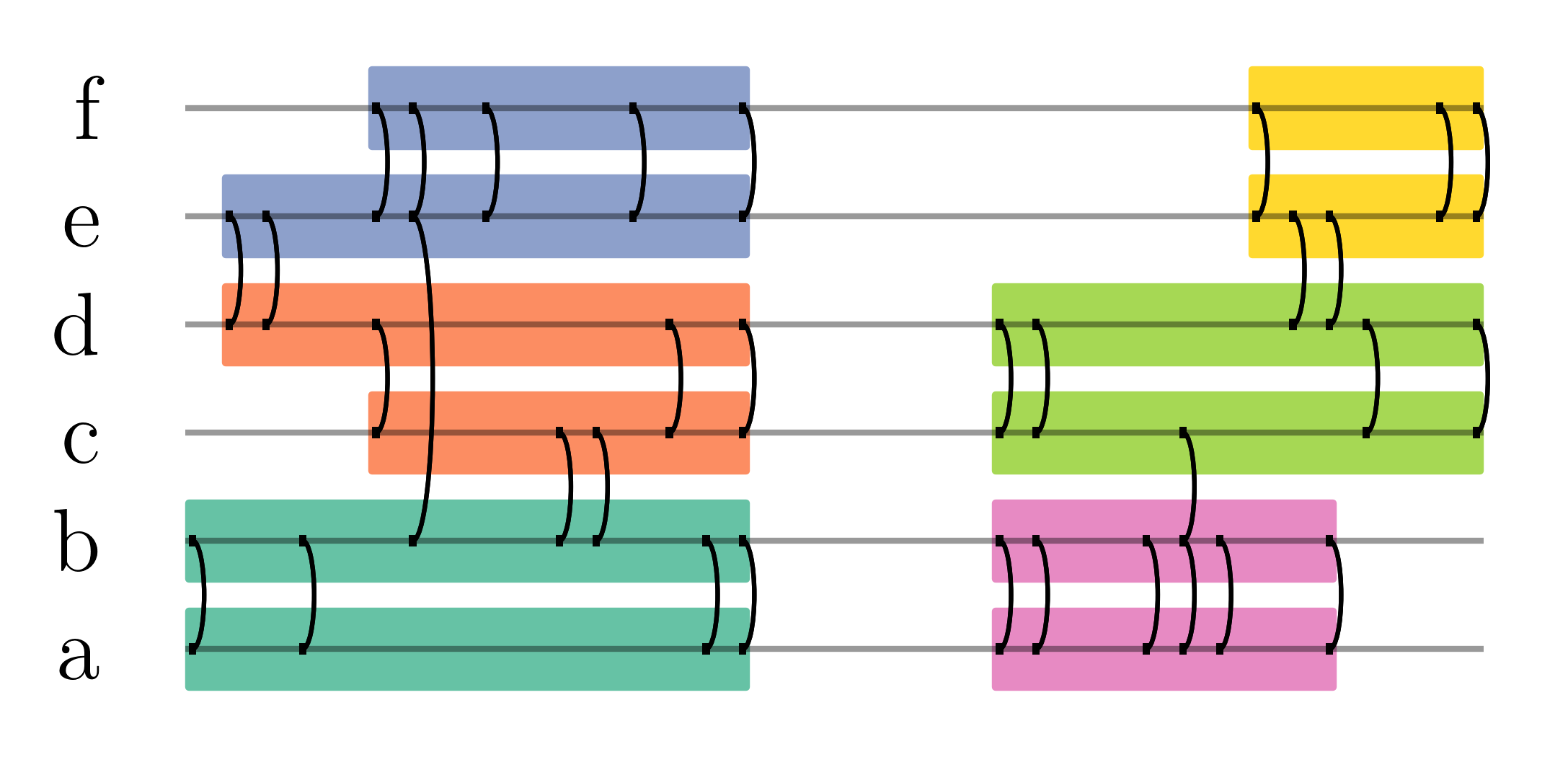}
        
        \vspace{2pt}
        \small (a) Favored by $(\gamma=2, \;\omega=0.5)$.
    \end{minipage}
    \hfill
    \begin{minipage}[b]{0.48\linewidth}
        \centering
        \includegraphics[width=\linewidth]{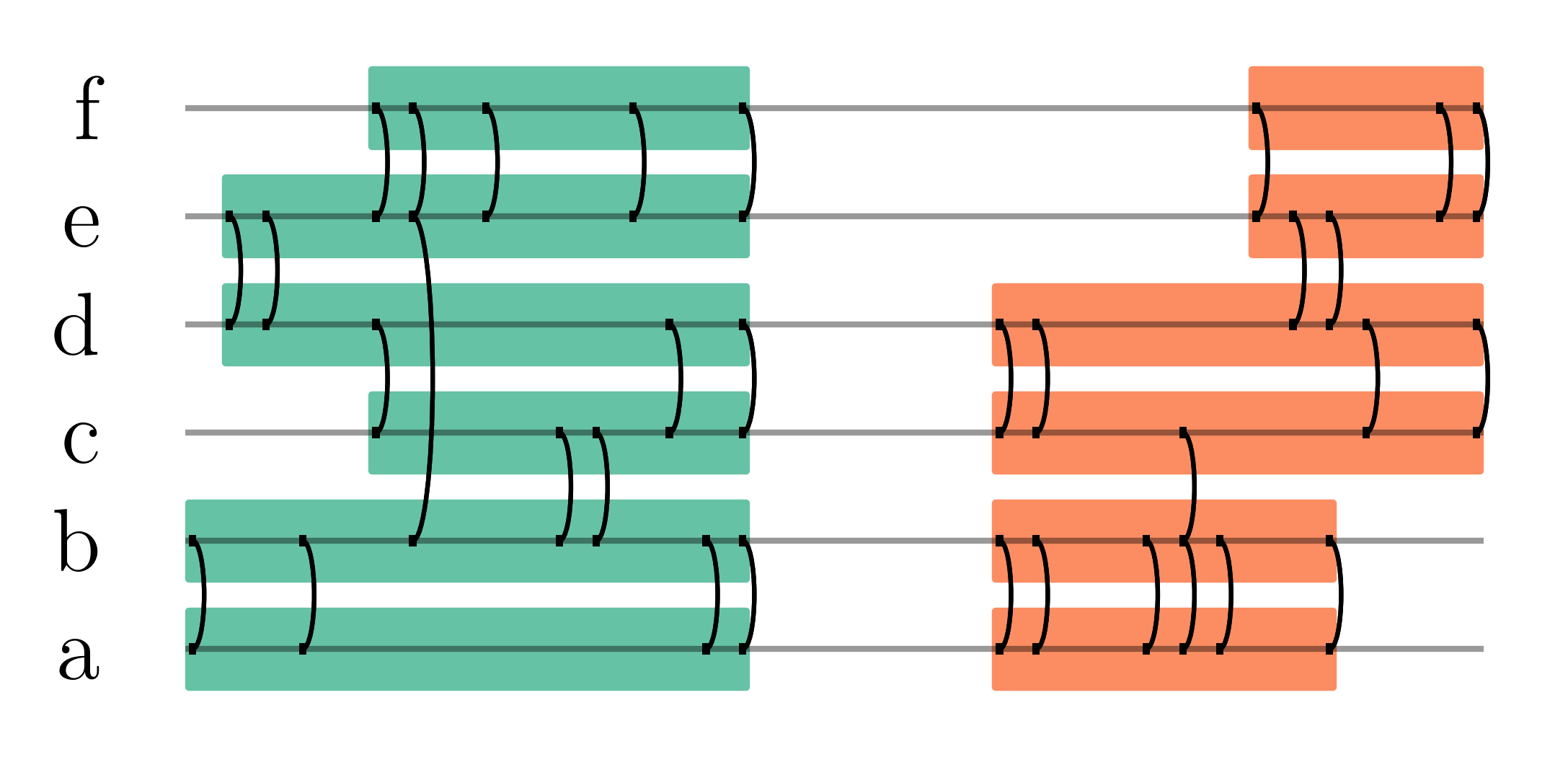}
        
        \vspace{2pt}
        \small (b) Favored by $(\gamma=0.25, \; \omega=0.5)$.
    \end{minipage}
    
    \vspace{3mm} 

    \begin{minipage}[b]{0.48\linewidth}
        \centering
        \includegraphics[width=\linewidth]{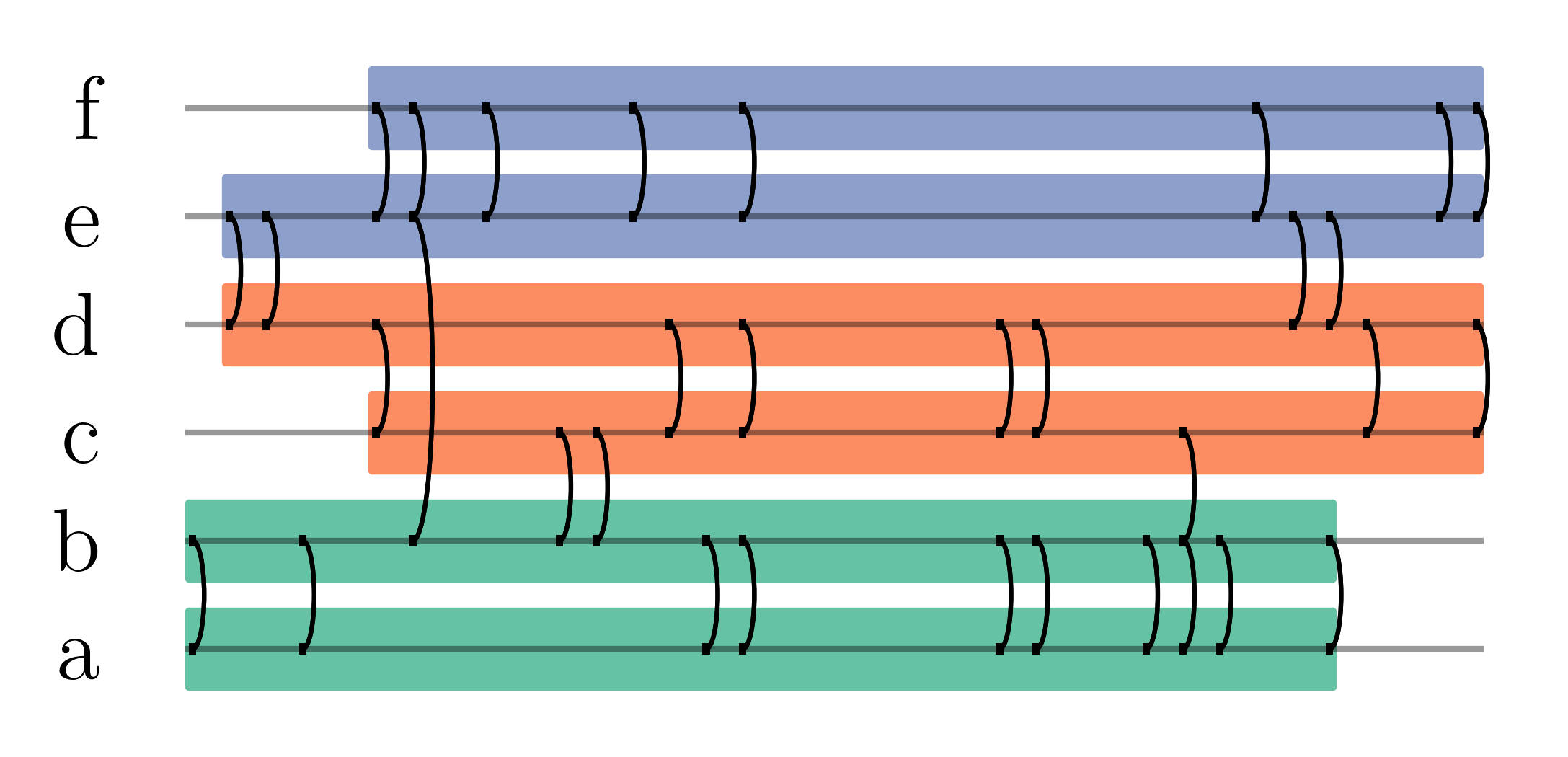}
        
        \vspace{2pt}
        \raggedright\small (c) Favored by $(\gamma=2, \; \omega=2)$.
    \end{minipage}
    \hfill
    \begin{minipage}[b]{0.48\linewidth}
        \centering
        \includegraphics[width=\linewidth]{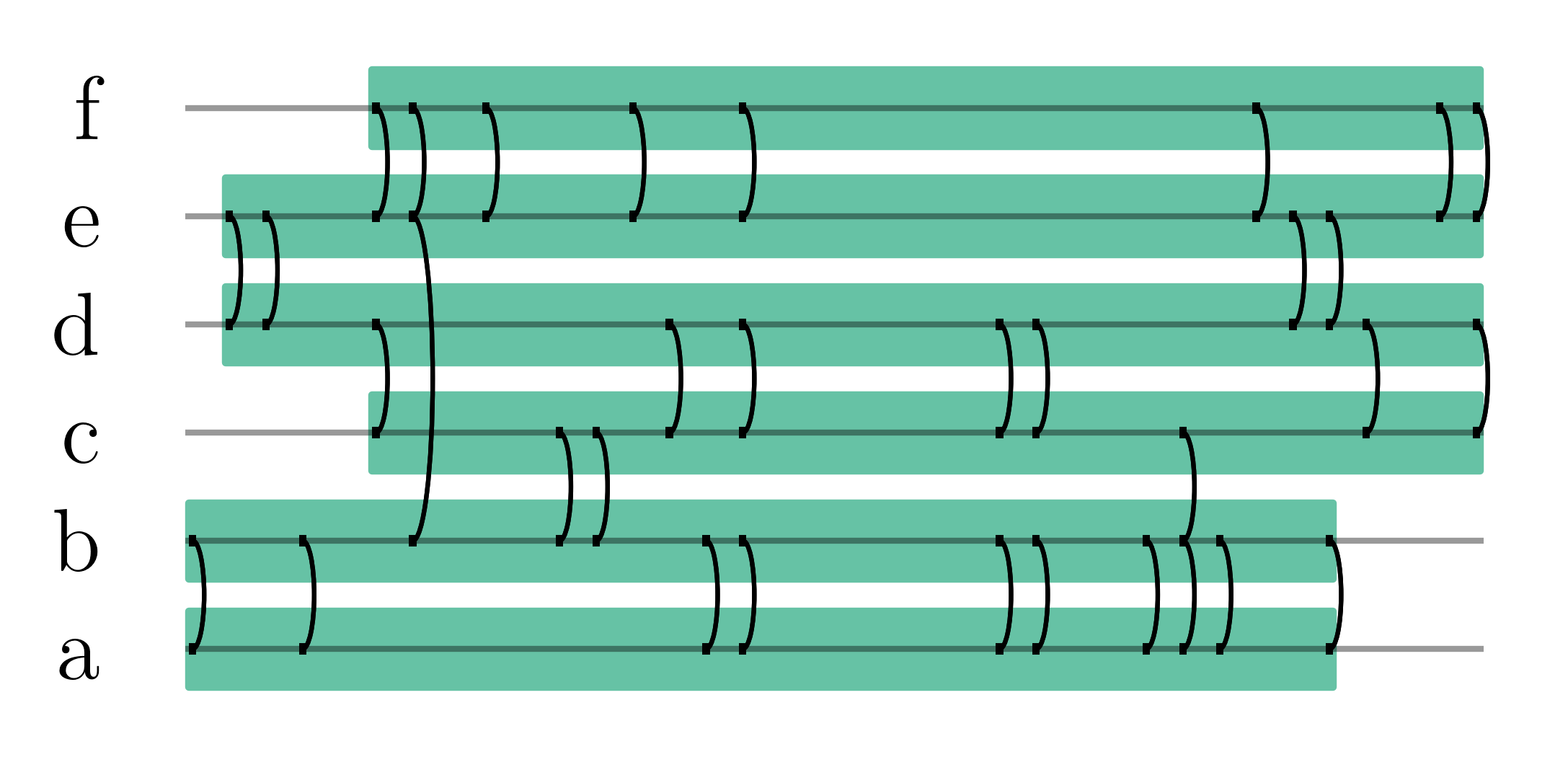}
        
        \vspace{2pt}
        \small (d) Favored by $(\gamma=0.25, \; \omega=2)$.
    \end{minipage}
    \vspace{4pt}
    \caption{\textbf{Impact of $\gamma$ and $\omega$ on temporal community structure.} Different community structures are shown in colors on a simple link stream, favored by different resolution parameters $\gamma$ and time smoothness terms $\omega$. Lower $\gamma$ favors fewer communities, and lower $\omega$ favors more changes over time.}
    \label{fig:resolutions}
\end{figure}

Figure~\ref{fig:resolutions} illustrates these effects on a simple instantaneous link stream. Panels show how different combinations of $(\gamma, \omega)$ influence the size and temporal smoothness of detected communities.

\subsubsection{Impacts of Interactions Types and Longitudinal Null Model}\label{sec:lex_vs_delayed}

The choice of temporal variant in the longitudinal null model directly affects the expected temporal shape of communities. 
The Joint-Membership (JM) expectation~\eqref{eq:le:jm} assumes that the overall community structure is stationary or nearly stationary: the expected amount of interaction is proportional to the total duration of the community, disregarding the specific times nodes are present in the community. 
In contrast, the Mean-Membership (MM) expectation~\eqref{eq:le:mm} accounts for the effective time its nodes spend in the community, allowing more flexible temporal patterns.

This distinction is particularly critical in link streams with delayed interactions. 
Under JM, a node may be considered part of a community as soon as an interaction \textit{targeted to it} begins, even if the node is not yet effectively engaged in the interaction. 
MM, on the other hand, only accounts for the time during which the node is actually involved, leading to different temporal community assignments.

\begin{figure}[htbp]
    \centering
    \begin{minipage}[b]{0.48\linewidth}
        \centering
        \includegraphics[width=\linewidth]{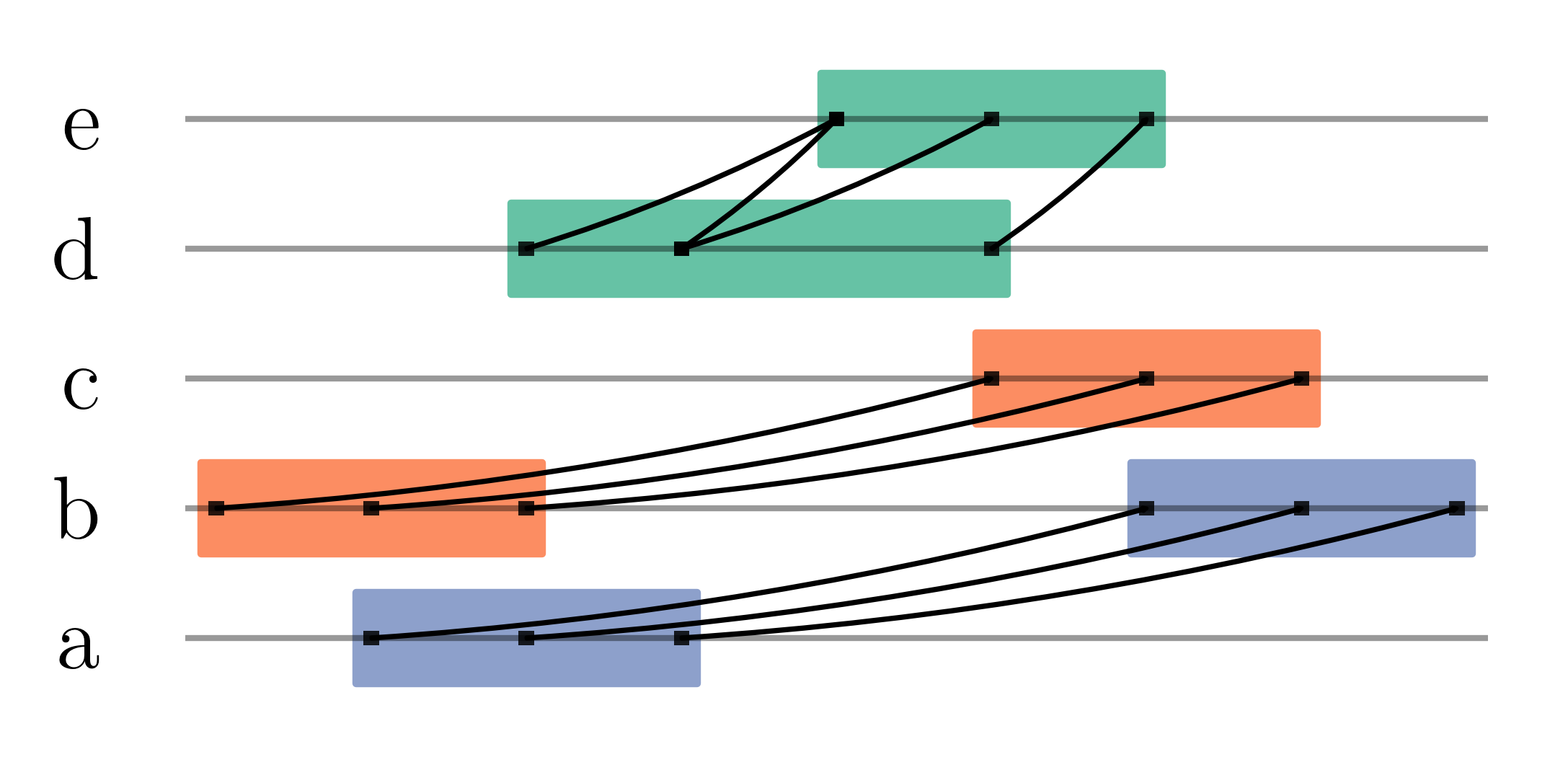}
        \vspace{2pt}
        \small (a) $Q_{\mathrm{MM}} = 0.75$, $Q_{\mathrm{JM}} = 0.6$
    \end{minipage}
    \hfill
    \begin{minipage}[b]{0.48\linewidth}
        \centering
        \includegraphics[width=\linewidth]{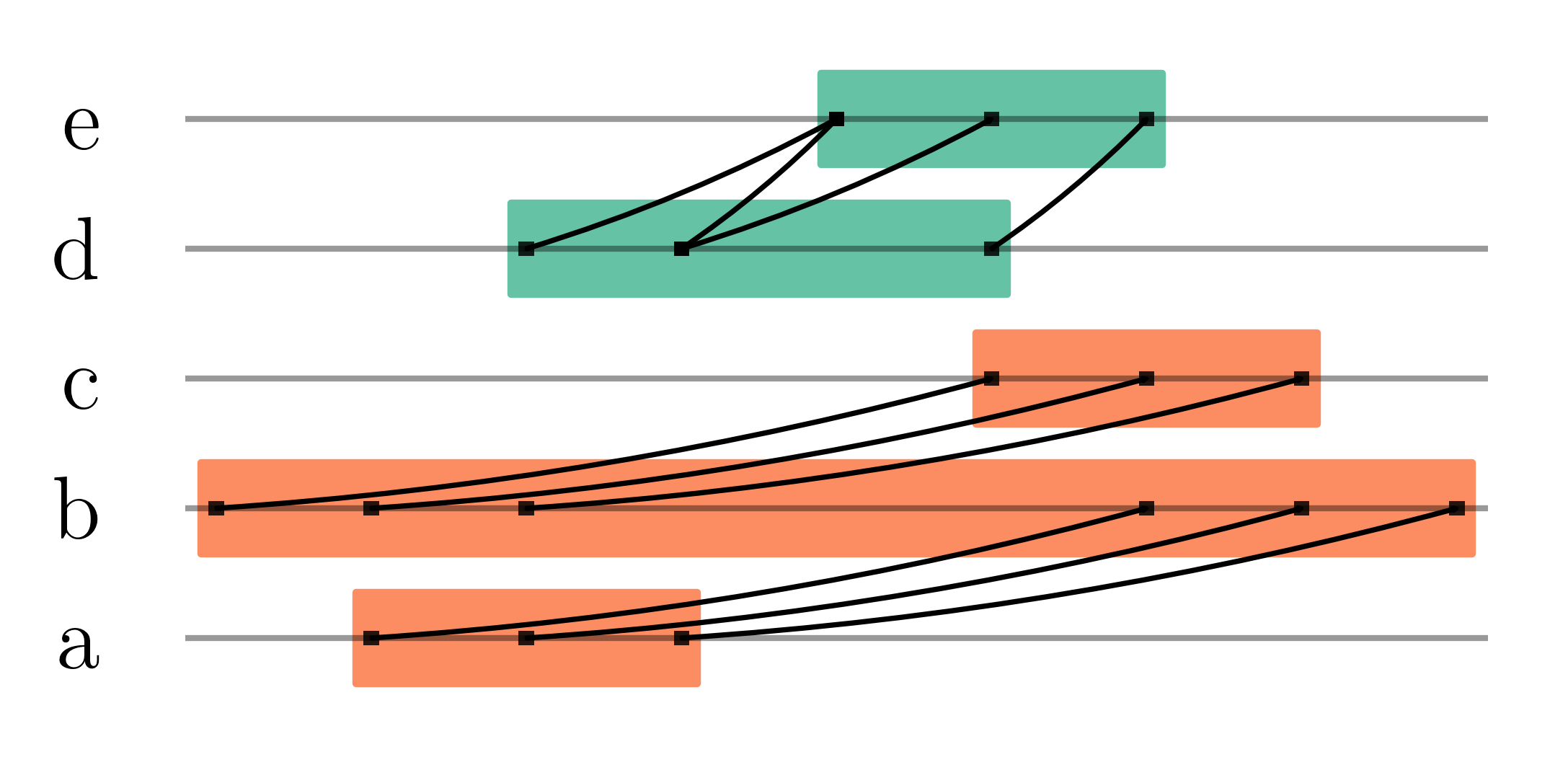}
        \vspace{2pt}
        \small (b) $Q_{\mathrm{MM}} = 0.71$, $Q_{\mathrm{JM}} = 0.65$
    \end{minipage}

    \vspace{3mm} 

    \begin{minipage}[b]{0.48\linewidth}
        \centering
        \includegraphics[width=\linewidth]{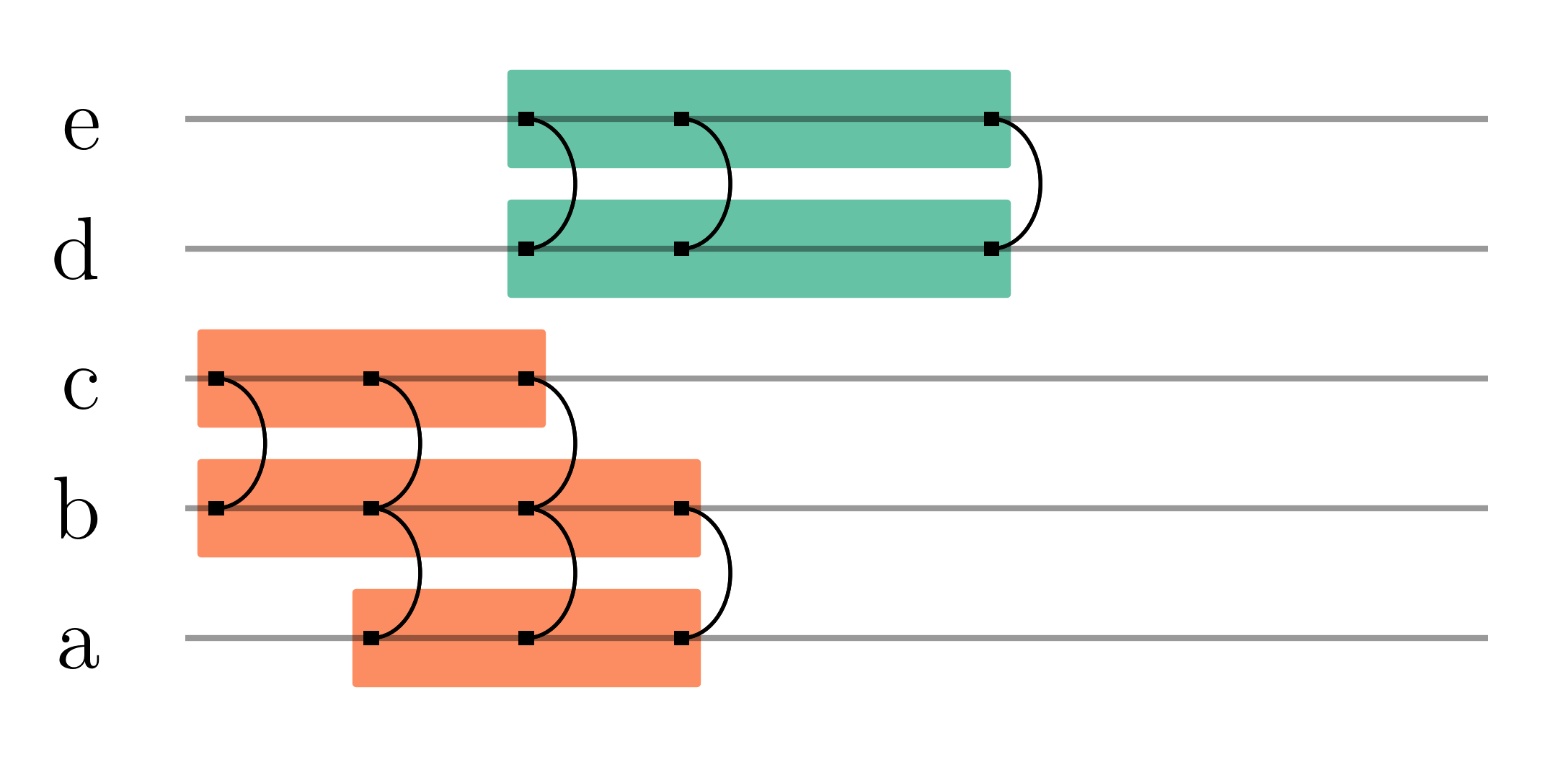}
        \vspace{2pt}
        \small (c) $Q_{\mathrm{MM}} = 0.78$, $Q_{\mathrm{JM}} = 0.75$
    \end{minipage}
    \hfill
    \begin{minipage}[b]{0.48\linewidth}
        \centering
        \includegraphics[width=\linewidth]{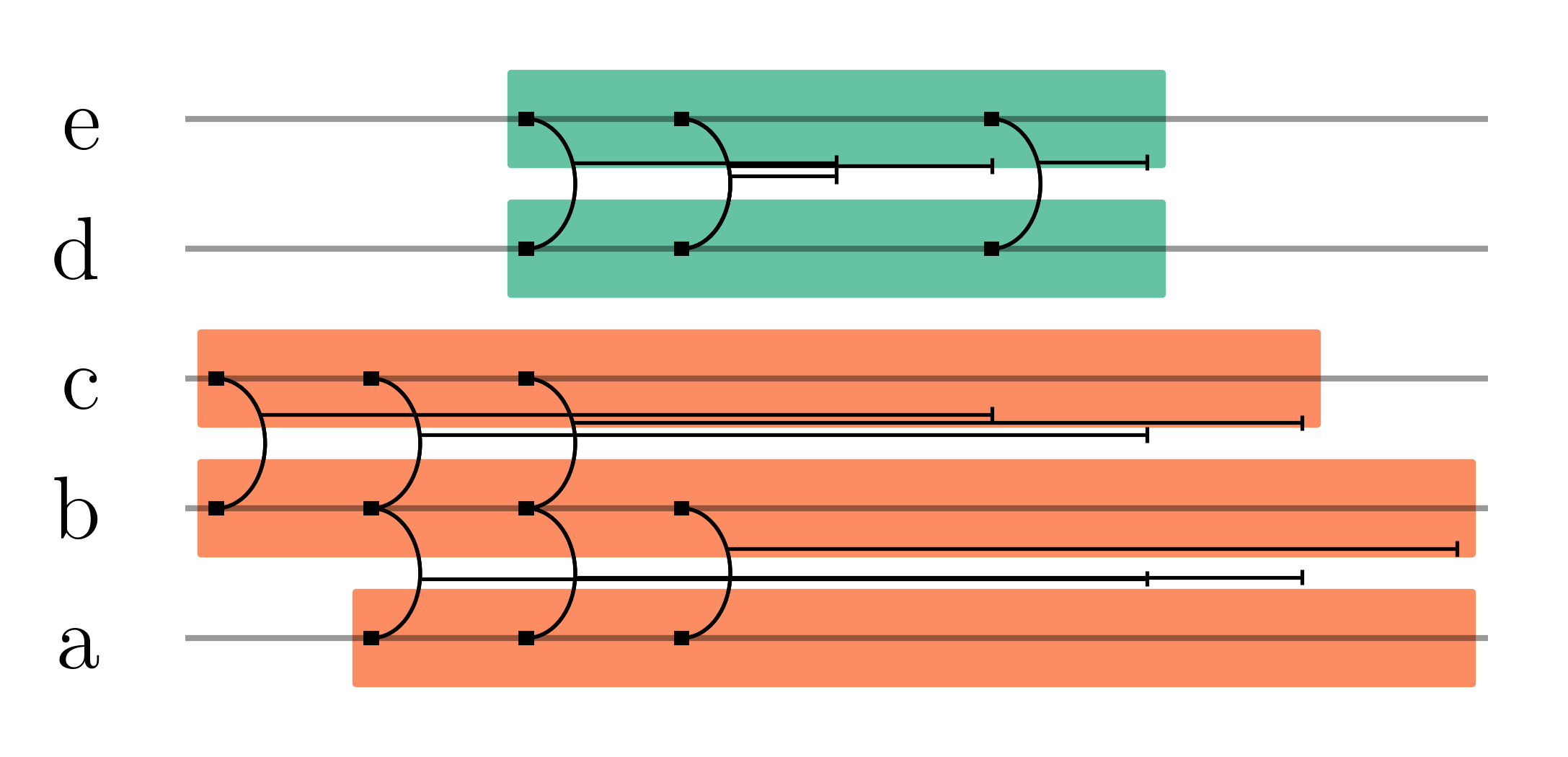}
        \vspace{2pt}
        \small (d) $Q_{\mathrm{MM}} = 0.52$, $Q_{\mathrm{JM}} = 0.49$
    \end{minipage}
    \vspace{4pt}
    \caption{\textbf{Four representations of a link stream with different interaction models and community structures.} 
    In panels (a) and (b), delayed interactions are preserved, whereas panels (c) and (d) represent alternative forms: instantaneous start times (c) and continuous intervals (d). Different community structures, displayed in colors, are evaluated using L-Modularity. Each structure is optimal with respect to $Q_{\mathrm{MM}}$, $Q_{\mathrm{JM}}$, or both. All evaluations use $\gamma = \omega = 1$. 
    Panels (a) and (b) are derived from the same temporal network but differ in community assignment; the MM model favors (a) over (b) ($0.75 > 0.71$), whereas the JM model favors (b) over (a) ($0.65 > 0.60$). This figure highlights the impact of interaction models and longitudinal null models on temporal community detection.}
    \label{fig:delayed_example}
\end{figure}

Figure~\ref{fig:delayed_example} illustrates these differences on a simple delayed link stream. 
In particular, (a) and (b) show two alternative community structures on the same temporal network: one better aligned with MM, favoring temporally localized memberships, and the other with JM, favoring more persistent communities. 
Although both partitions are evaluated using the same quality function (L-Modularity), their scores differ depending on the null model: MM favors (a), whereas JM favors (b). 

Additionally, (c) and (d) illustrate on a toy example how different modeling choices for delayed interactions (instantaneous vs.\ interval-based representations) can alter the temporal structure of the network. 
These representations reduce the discrepancy between JM and MM, yielding community structures that are simultaneously favored by both null models. 

Overall, this example highlights that both the interaction model and the longitudinal null model jointly shape the detected temporal communities, with particularly strong effects in the presence of delayed interactions, emphasizing the importance of accurately modeling real-world data.

\section{Dynamic Community Detection: LAGO}\label{sec:lago}

LAGO (Longitudinal Agglomerative Greedy Optimization) has been introduced in~\cite{brabant2025discoveringcommunitiescontinuoustimetemporal} to optimize L-Modularity on simple instantaneous link streams, and thereby uncover dynamic communities in temporal networks.
These communities correspond to so-called \emph{time modules}~\cite{brabant2025discoveringcommunitiescontinuoustimetemporal}, i.e., sub link streams that exhibit both temporal and topological coherence with respect to their contribution to the L-Modularity score.
Formally, the optimization procedure produces time modules, which are then interpreted as dynamic communities.

\subsection{LAGO For Simple Link Streams}

LAGO builds upon the Louvain method~\cite{Blondel_2008} (see Section~\ref{sec:rel:stat:louv}) and incorporates ideas inspired by Infomap~\cite{Rosvall_2009} and Leiden~\cite{Traag_2019}, together with components specifically designed for link streams. The method was introduced to optimize L-Modularity on simple instantaneous link streams. Without discussing all the details of the algorithm (available in the original article), we summarize here the key elements to understand the adaptations required to optimize the generalized L-Modularity. 

LAGO is composed of a \textit{core loop}, which proceeds in two alternating phases: i) time module reassignment and ii) aggregation. The reassignment concerns atomic elements called \textit{active time nodes}, that play a role equivalent to nodes in the original Louvain. At each step, the algorithm performs a \textit{local move}, in which the active time nodes test the interest of joining the community of one of their \textit{neighbors in the link stream}. If the move is accepted, all time segments delimited by consecutive temporal neighbors involved in both the source and target communities are reassigned accordingly.

While the core loop remains unchanged, we detail here the changes required to the definition of active time nodes and neighborhoods in link streams. We also detail the changes required to the \textit{variants} of LAGO introduced in the original article. An illustration of the modified core loops can be found in Fig. \ref{fig:lago:discrete:example} (timestamp-based) and Fig. \ref{fig:lago:continuous} (interval-based).

\subsection{LAGO for Generalized L-Modularity}

In this section, we detail how LAGO can be adapted to optimize the generalized L-modularity introduced in Definition~\ref{generalizedLmodularity}. 

\subsubsection{Active Time Nodes}

In the original LAGO, the authors introduce the concept of \textit{active time nodes}, which play the role of the atomic elements manipulated by the algorithm, equivalent to nodes in the static Louvain.
\begin{definition}\label{def:activetimenodes}
    Let $L = (T, V, E)$ be a simple instantaneous link stream. The set $\mathcal{A}$ of \textbf{active time nodes} is defined as the set of node--time pairs involved in interactions:
    \begin{align}
         \mathcal{A} = \left\{ (u,t) \in V \times T \;\middle|\; \exists \; v \in V \text{ such that } (u, v, t) \in E\right \}
    \end{align}
\end{definition}

In generalized LAGO, the definition of active time nodes must be modified to account for both incoming and outgoing timestamp-based interactions in:
\begin{definition}\label{def:activetimenodes:dir}
    Let $L = (T, V, E)$ be a general timestamp-based link stream. The set $\mathcal{A}$ of \textbf{active time nodes} is defined as the set of node--time pairs involved in interactions:
    \begin{align}
         \mathcal{A} = \left\{ (u,t) \in V \times T \;\middle|\; \exists \; v \in V, t' \in T \text{ such that } (u,v,t,t', \cdot) \in E \text{ or } (v, u,t',t, \cdot) \in E \right \}.
    \end{align}
\end{definition}
This definition also applies without modification to \textbf{weighted}, \textbf{delayed}, and \textbf{multipartite} interactions.


For \textbf{interval-based} interactions, the notion of active time nodes is no longer suitable.  
Instead, we introduce \emph{active time-segment nodes}. 
In the continuous setting, the atomic elements manipulated by the algorithm must correspond to temporal units that can be uniquely assigned to a time module. Directly using interaction intervals as atomic elements is problematic because overlapping intervals cannot be unambiguously assigned to a single time module. 
To avoid this issue, we redefine the set of interval-based interactions by segmenting them so that no interaction starts or ends within any resulting segment. This ensures that the resulting segments form a partition of time where interaction boundaries are aligned.
Formally, each interaction interval is \emph{segmented} so that no other interaction starts or ends inside the resulting time segments. The resulting segments are then used to define the atomic elements manipulated by the algorithm, referred to as \emph{active time-segment nodes}. Let
 \[
 \mathcal{T} = \bigcup_{u,v,t_s, t_e \in E} \{ t_s, t_e \}
 \] 
 be the set of all start and end times of interaction intervals, and let $\mathcal{S}([t_s, t_e))$ denote the partition of $[t_s, t_e)$ induced by $\mathcal{T}$, defined by $t_s = t_0 < t_1 < \dots < t_{n+1} = t_e$, that is $\mathcal{S} \left([t_s, t_e) \right) = \left \{ [t_i, t_{i+1}) \right \}_{i=0}^{n}$. We denote
 \[
 E_{\mathcal{S}} = \bigcup_{uvt_s t_e \in E} \mathcal{S}([t_s, t_e))
 \]
 the set of continuous interactions after \textit{segmentation}. 
 LAGO can then be applied on \emph{active time-segment nodes}.
 \begin{definition}\label{def:activetimesegnodes}
    Let $L = (T, V, E)$ be a general link stream with interval-based interactions. The set $\mathcal{A}_{\mathcal{S}}$ of \textbf{active time-segment nodes} is defined as the set of node--time-segment pairs involved in interactions:
    \begin{align}
         \mathcal{A}_{\mathcal{S}} = \left\{ (u, [t_s, t_e)) \mid \exists \; v \in V \text{ such that } (u, v, t_s, t_e, \cdot) \in E_{\mathcal{S}} \text{ or } (v, u, t_s, t_e, \cdot) \in E_{\mathcal{S}} \right\}.
    \end{align}
\end{definition}
As illustrated in Figure~\ref{fig:lago:continuous} on an interval-based link stream, LAGO initially assigns each active time-segment node to its own time module, and then moves them between neighbor time modules. The move phase does not impact other active time-segment node affiliations since the segmented set of interval-based interactions prevents overlapping by design. 

The \emph{trimmed community property} demonstrated
in~\cite{brabant2025discoveringcommunitiescontinuoustimetemporal}, which
justifies restricting the optimization of L-Modularity to active time nodes in the instantaneous setting, naturally extends to the continuous case when considering active time-segment nodes. Extending dynamic communities towards non-active time segments lowers L-Modularity, analogously to the instantaneous case.

Note that considering $E_{\mathcal{S}}$ instead of $E$ changes neither the weight of the network $w$, nor the weight of interactions $W_{uv \in C}$ from $u$ to $v$ within community $C$, nor the in or out degree $k_{u}^{\mathrm{in}}, k_{u}^{\mathrm{out}}$ of a node $u$. Only the number $m$ of interactions is affected. LAGO is applied with the initial number of interactions $m$, computed before the segmentation phase.

\subsubsection{Neighborhood in Link Streams} 

In the original LAGO, the notion of neighbors is redefined as a combination of temporal and topological neighbors, so that the algorithm tests extensions of communities in both modes. The definition is slightly adapted in the generalized case as follows:

\textit{Topological neighbors} correspond, as in static networks, to nodes involved in the same interaction. In the directed setting, the direction of the interaction is ignored when defining neighbors, while for delayed interactions, the corresponding active time nodes may occur at different timestamps.

\textit{Temporal neighbors} correspond to the previous and next active occurrences of the same node. For interval-based interactions, the same definition extends to active time-segment nodes.

The segmentation procedure introduced above ensures that, in the interval-based setting, interacting nodes share time segments of identical duration, which guarantees well-defined topological neighbors.

\subsubsection{Local Moves} 
\begin{figure}[htbp]
    \centering
    \begin{minipage}[b]{0.48\linewidth}
        \centering
        \includegraphics[width=\linewidth]{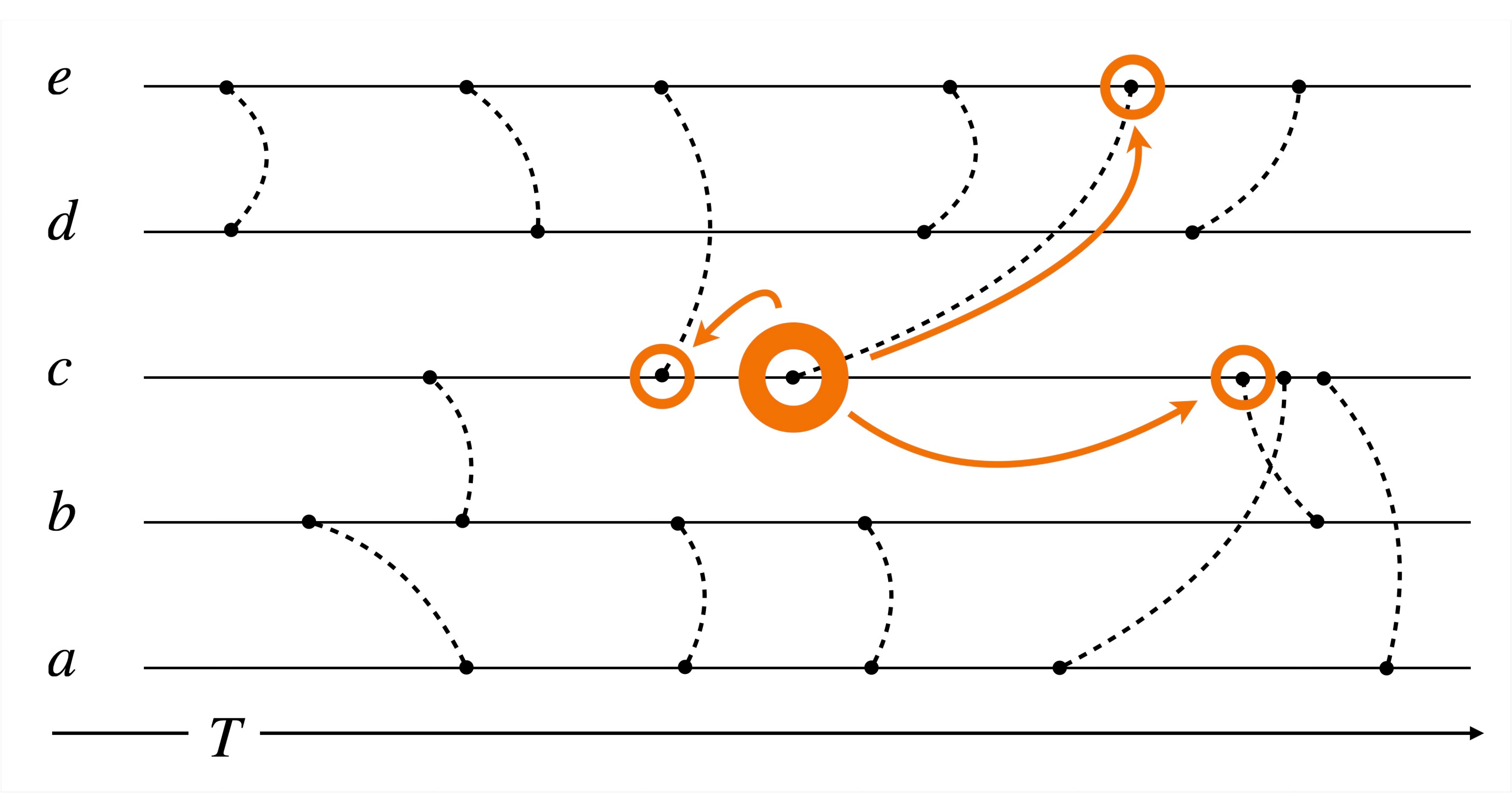}
        \vspace{2pt}
        \small (a) Exploration phase.
    \end{minipage}
    \hfill
    \begin{minipage}[b]{0.48\linewidth}
        \centering
        \includegraphics[width=\linewidth]{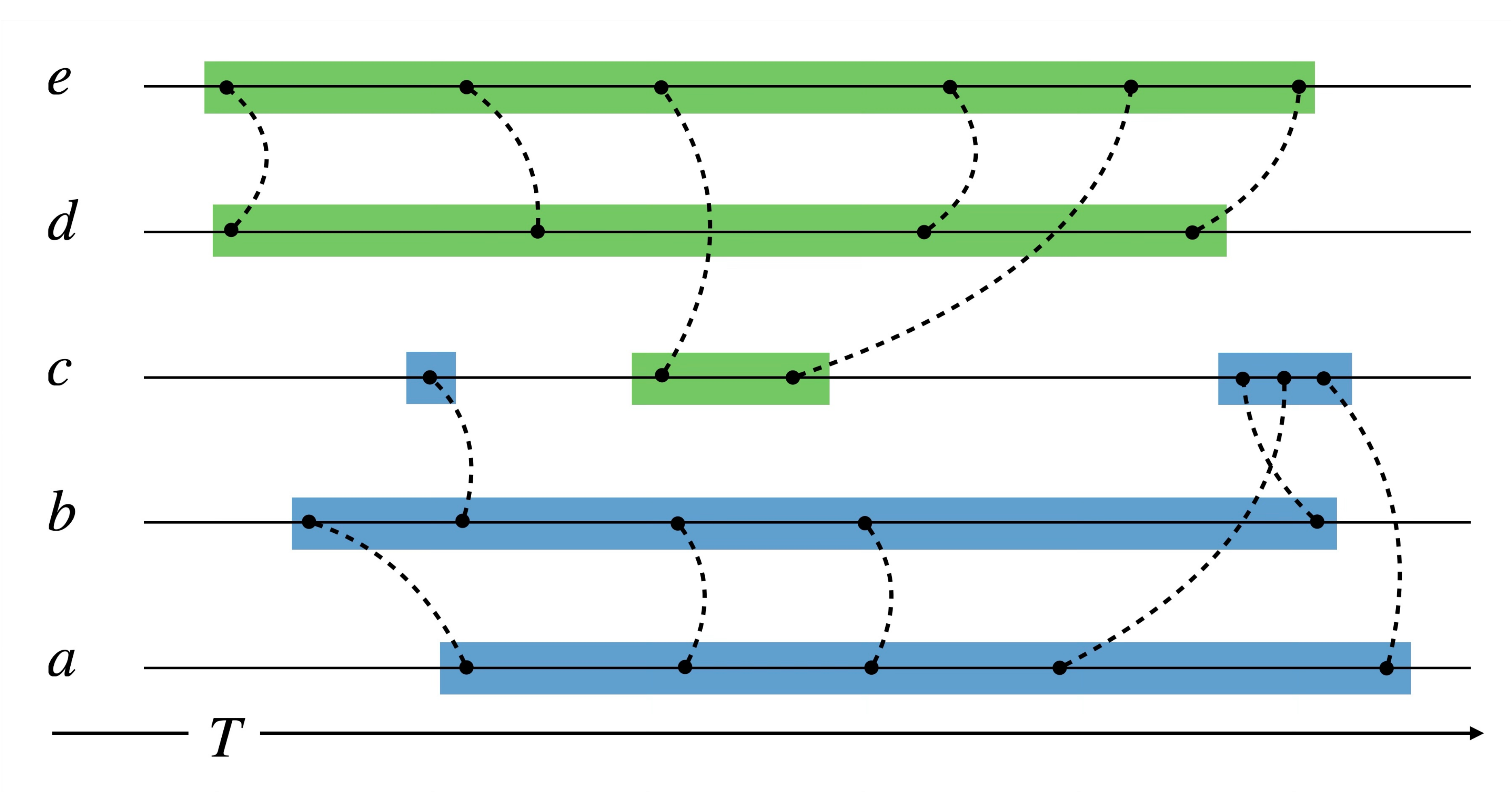}
        \vspace{2pt}
        \small (b) Result of LAGO.
    \end{minipage}

    \vspace{4pt}
    \caption{\textbf{Example of LAGO application on a timestamp-based link stream.} Each \emph{active time node} (black dots) initially belongs to its own time module. Then, in panel (a), each one (e.g., thick orange-circled) explores among its \textit{neighbors} (thin orange-circled) the best move for reassignment. Once no move improves L-Modularity anymore, LAGO stops, yielding two dynamic communities in blue and green (panel b). The time segments between successive active time nodes are assigned to a community when both endpoints belong to the same community; otherwise they remain unassigned, highlighting transitions in node membership over time.}
    \label{fig:lago:discrete:example}
\end{figure}

\begin{figure}[htbp]
    \centering
    \begin{minipage}[b]{0.48\linewidth}
        \centering
        \includegraphics[width=\linewidth]{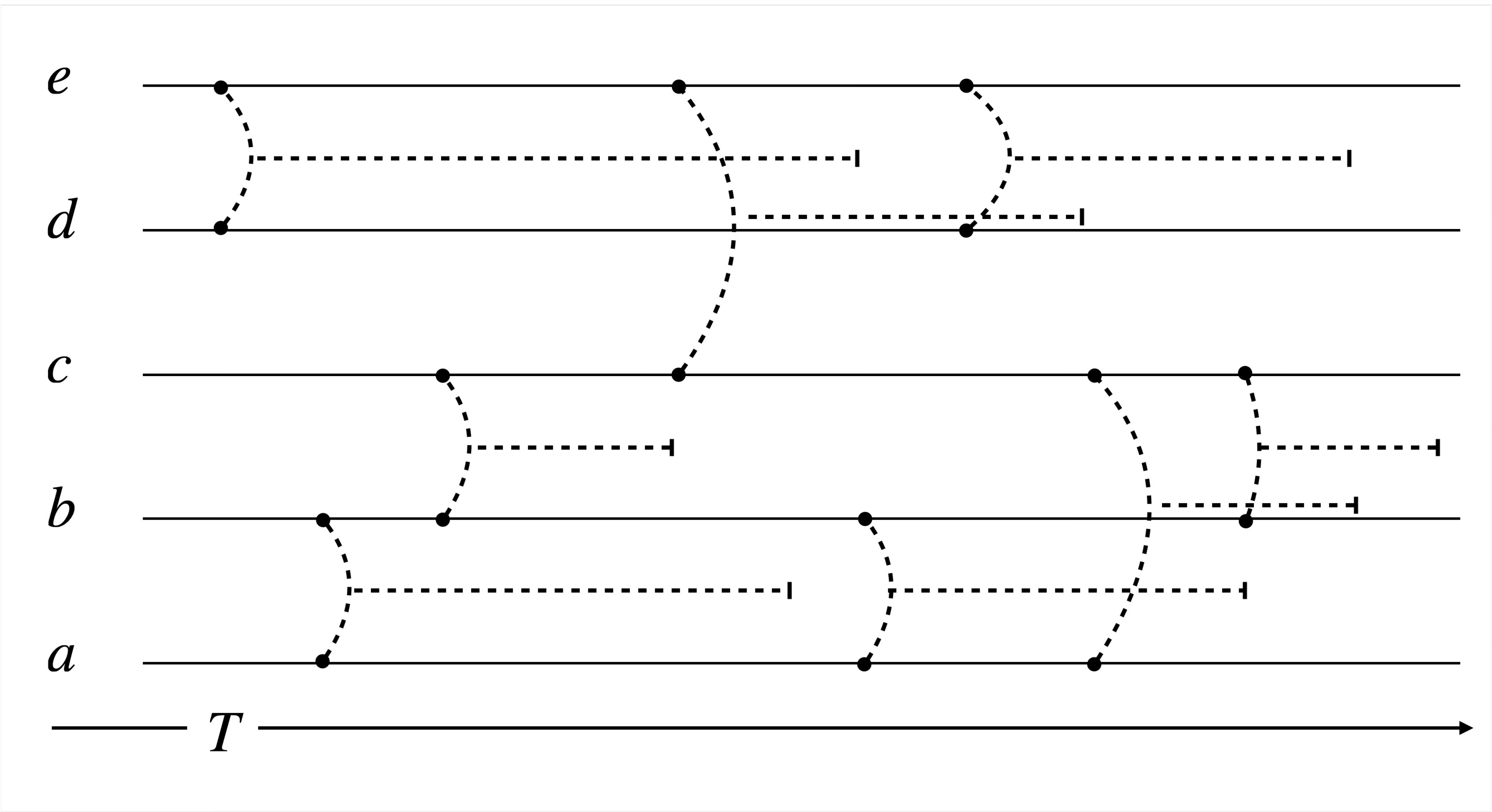}
        \vspace{2pt}
        \small (a) Initial interval-based link stream.
    \end{minipage}
    \hfill
    \begin{minipage}[b]{0.48\linewidth}
        \centering
        \includegraphics[width=\linewidth]{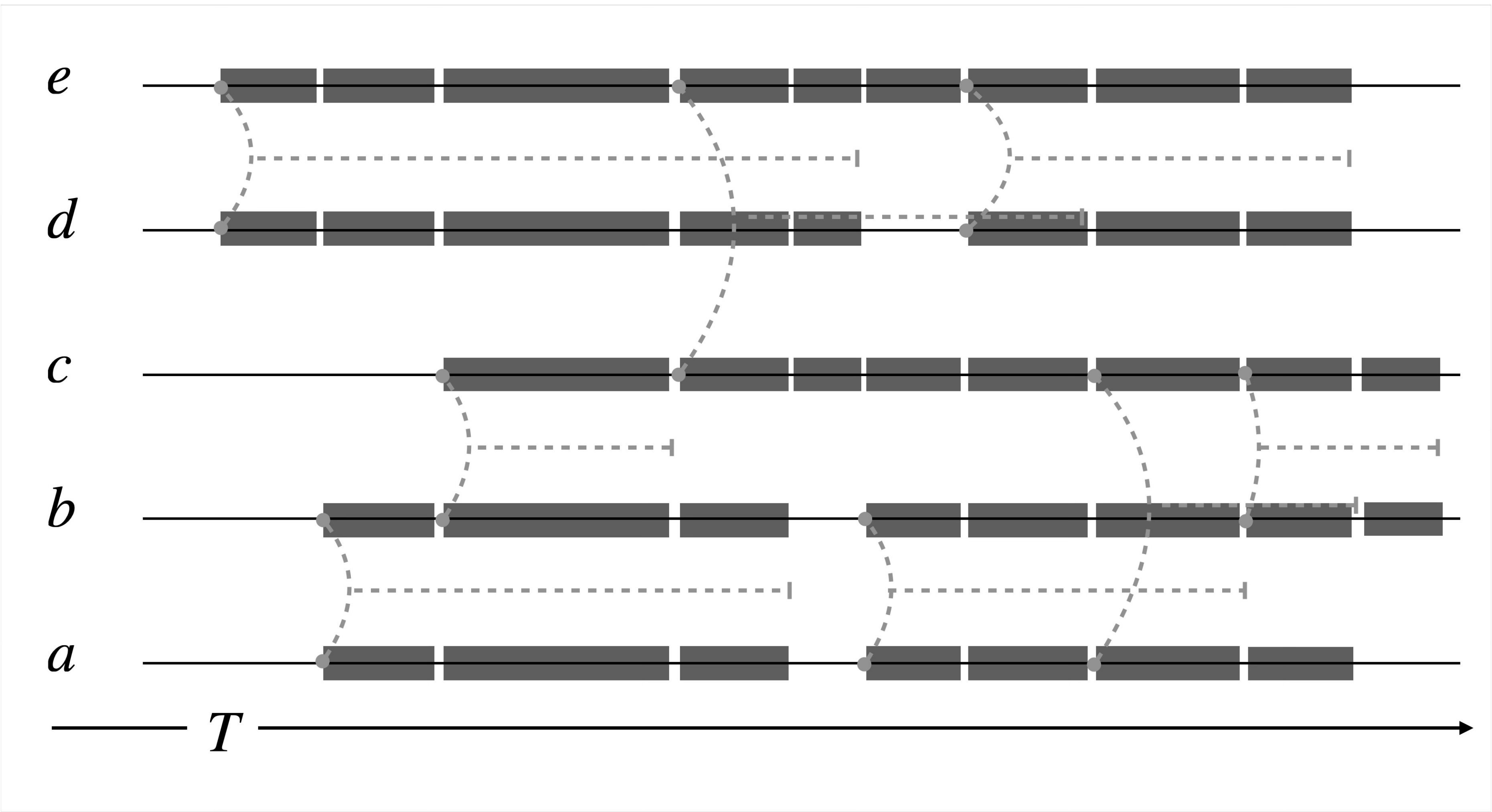}
        \vspace{2pt}
        \small (b) \emph{Segmentation} phase.
    \end{minipage}
    
    \vspace{3mm} 

    \begin{minipage}[b]{0.48\linewidth}
        \centering
        \includegraphics[width=\linewidth]{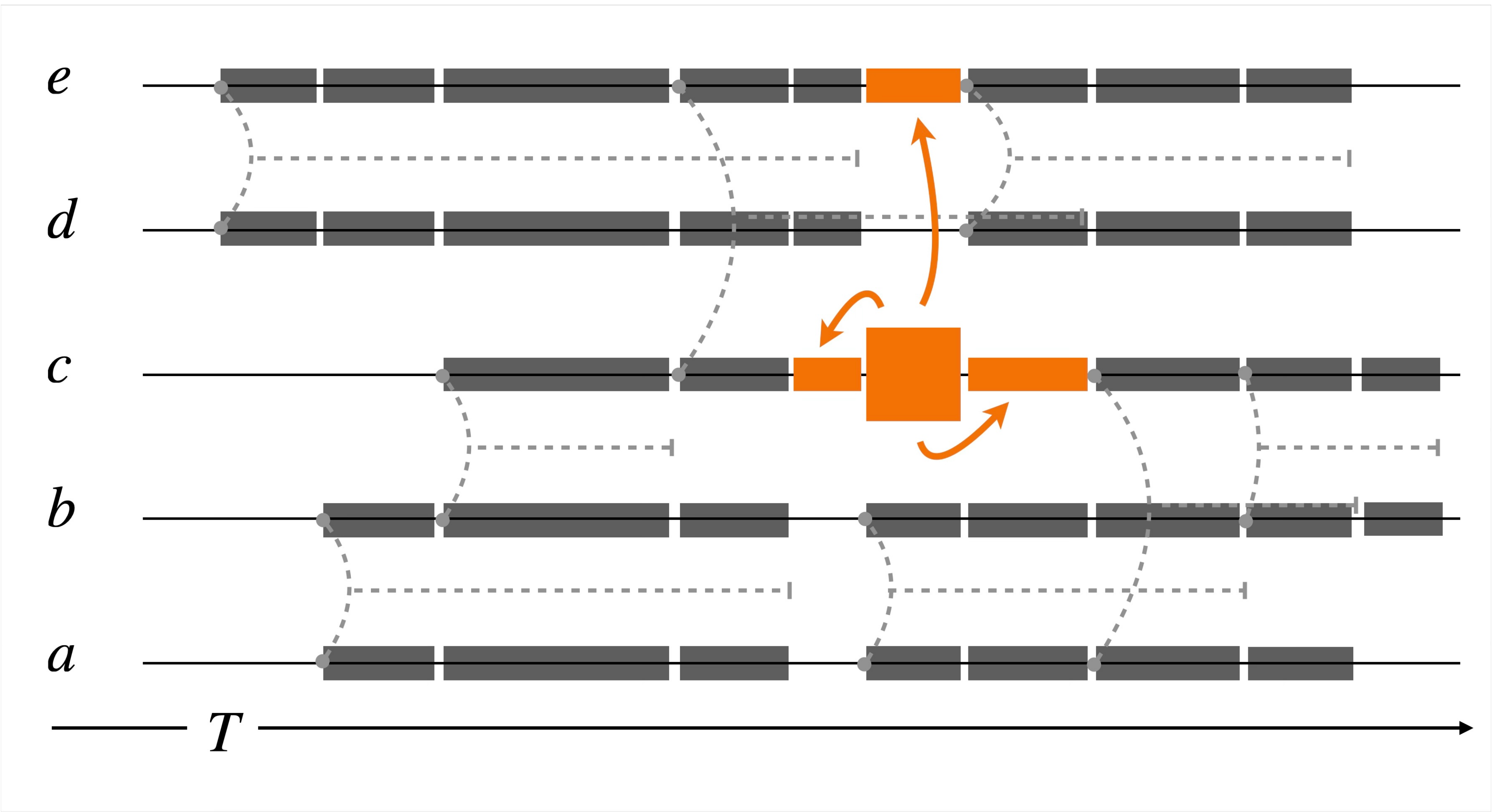}
        \vspace{2pt}
        \small (c) LAGO exploration.
    \end{minipage}
    \hfill
    \begin{minipage}[b]{0.48\linewidth}
        \centering
        \includegraphics[width=\linewidth]{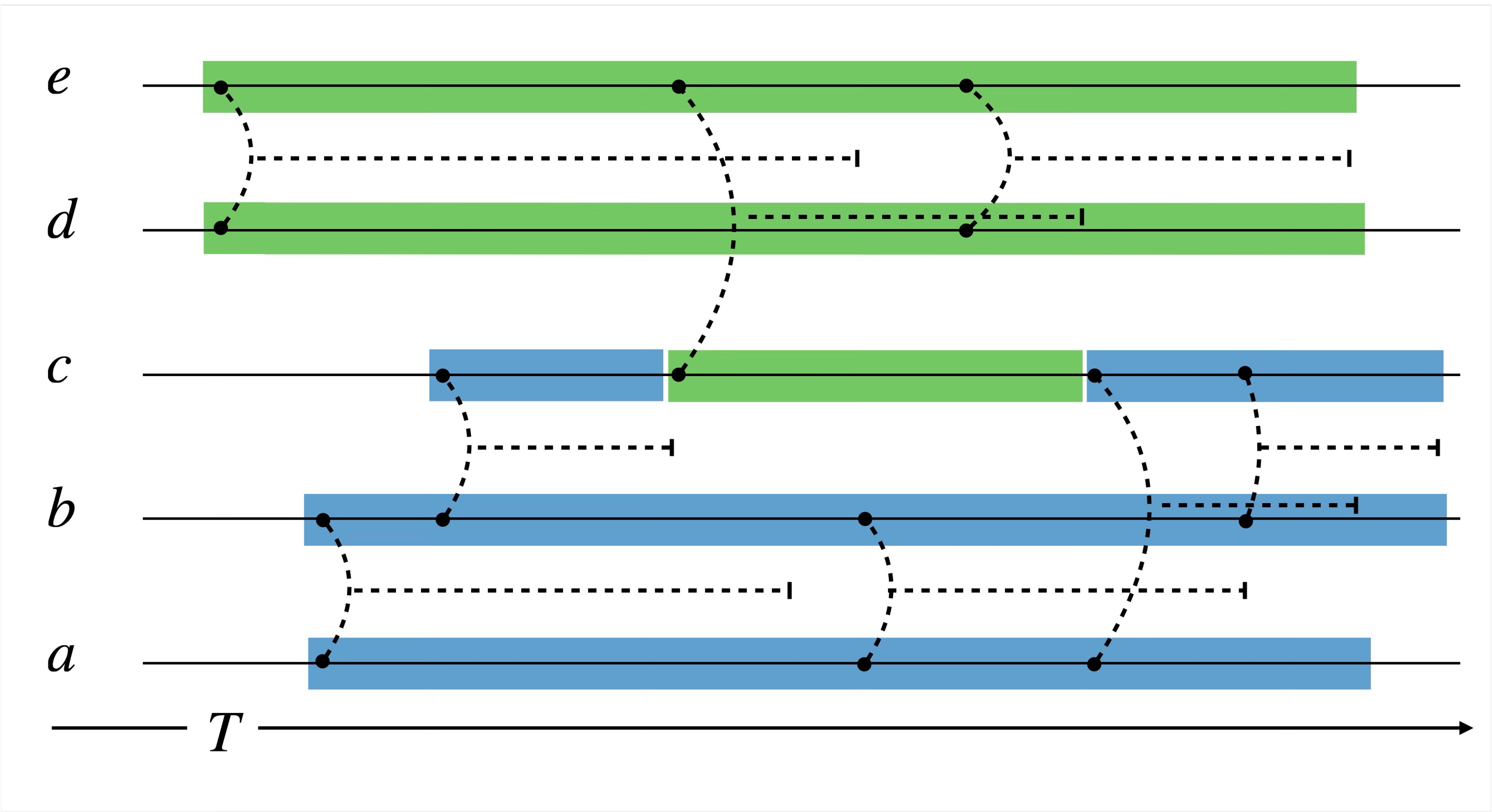}
        \vspace{2pt}
        \small (d) Result of LAGO.
    \end{minipage}

    \vspace{4pt} 
    \caption{\textbf{Example of LAGO application on an interval-based link stream.} 
    The interval-based link stream (panel a) is first \emph{segmented} so each \emph{active time-segment node} (black rectangles) has at most one neighbor per node in time (panel b). Each active time-segment node initially belongs to its own time module. Then, in panel (c), each one (e.g., thick orange segment) explores among its \textit{neighbors} (thin orange segments) the best move for reassignment. Once no move improves L-Modularity, LAGO stops, yielding two dynamic communities in blue and green (panel d).}
    \label{fig:lago:continuous}
\end{figure}

In LAGO, the \textit{local move} procedure remains unchanged in the generalized setting. 
For interval-based interactions, the reassignment rule naturally extends to \textit{active time-segment nodes}: when a move is accepted, all time segments delimited by consecutive temporal neighbors and involved in both the source and target communities are reassigned accordingly. Figures~\ref{fig:lago:discrete:example}, \ref{fig:lago:continuous} illustrate this mechanism for both timestamp-based and interval-based interactions.

\subsubsection{Variations} The original LAGO includes several variations. Among them, STNM and STEM require adaptation for interval-based interactions: they operate on active time-segment nodes instead of active time nodes, with the segmentation procedure ensuring that moves do not create inconsistencies with overlapping time intervals in time modules.
All other variants, including fast exploration mode and RIR, remain directly applicable without modification.

\section{Applications on Real-World Data}\label{sec:applications}

In this section, we illustrate the application of our contribution for dynamic community detection beyond simple temporal networks on real-world data. We demonstrate that LAGO successfully optimizes L-Modularity on those real-world datasets, and that the communities discovered provide reasonable insights.
We illustrate results on weighted and directed temporal networks in Section~\ref{sec:eurovision}; interval-based and delayed interactions in Section~\ref{sec:citibike}; and bipartite temporal networks in Section~\ref{sec:bluesky}. Table~\ref{tab:temporal_networks} summarizes core features of the mentioned networks.

\begin{table}[h]
\centering
\caption{
\textbf{Summary of the real-world temporal networks.}
All datasets are preprocessed to reduce their volume and facilitate visualization.
The table reports the main structural and temporal properties of each network after preprocessing.
\textit{\# Active times} corresponds to the number of distinct time instants with at least one recorded interaction.
\textit{\# Active time nodes} corresponds to the number of distinct node--time pairs $(u, t)$ (or node--time-segment pairs for interval-based interactions) participating in at least one interaction (see Definitions~\ref{def:activetimenodes:dir}, \ref{def:activetimesegnodes}).
Averages are computed over interactions.}\label{tab:temporal_networks}
\begin{tabular}{@{}l*{7}{c}@{}}
\toprule
 & \multicolumn{4}{c}{Eurovision (\ref{sec:eurovision})} & \multicolumn{2}{c}{Citi Bike (\ref{sec:citibike})} & Bluesky (\ref{sec:bluesky})\\
\cmidrule(lr){2-5} \cmidrule(lr){6-7} \cmidrule(lr){8-8}
 & \ref{fig:eurovision}.a 
 & \ref{fig:eurovision}.b 
 & \ref{fig:eurovision}.c 
 & \ref{fig:eurovision}.d 
 & \ref{fig:citibike}.a 
 & \ref{fig:citibike}.b 
 & \ref{fig:bluesky} \\
\midrule
Duration & \multicolumn{4}{c}{21 years} & \multicolumn{2}{c}{4 hours} & 9 months \\
\# Nodes & \multicolumn{4}{c}{49} & \multicolumn{2}{c}{46} & 283 + 259 \\
\# Active times & \multicolumn{4}{c}{21} & -- & 176 & 3 598 \\
\# Active time nodes & \multicolumn{4}{c}{696} & 413 & 411 & 16 550 \\
\# Interactions & 3 224 & 3 224 & 2 816 & 2 816 & \multicolumn{2}{c}{214} & 10 179 \\
Total weight & 8 487 & -- & 8 487 & -- & \multicolumn{2}{c}{--} & -- \\
Avg weight & 2.63 & -- & 3 & -- & \multicolumn{2}{c}{--} & -- \\
Avg duration / delay & -- & -- & -- & -- & \multicolumn{2}{c}{23 minutes}   & -- \\
\midrule
Instantaneous & x & x & x & x & -- & -- & x \\
Delayed & -- & -- & -- & -- & -- & x & -- \\
Interval-based & -- & -- & -- & -- & x & -- & -- \\
\midrule
Weighted & x & -- & x & -- &  --& -- & -- \\
Directed & x & x & -- & -- & -- &  --& -- \\
Multipartite &  --&  --& -- & -- & -- &  --& x \\
\botrule
\end{tabular}
\end{table}

\subsection{A Voting Network: the Eurovision Song Contest}\label{sec:eurovision}

We apply LAGO to the voting network of the Eurovision Song Contest, an annual international music competition in which participating countries submit original songs. Under the pre-2016 voting scheme, each country assigns a single set of points (1--8, 10, and 12) to its top ten performances, based on an aggregation of jury and televote rankings.

Several studies have analyzed Eurovision data from a network perspective, highlighting voting biases that cannot be explained solely by song quality \cite{melodyeurope,preference_mantzaris_2018,Garcia2013-fk,Dekker2007}. Without aiming to revisit these findings in depth, we model the data as a temporal network and apply LAGO to investigate whether such voting patterns naturally emerge.

The dataset\footnote{\url{https://github.com/Spijkervet/eurovision-dataset/releases/tag/2023}} contains detailed voting records, including the voting country, the receiving country, the year, and the number of points awarded. These data naturally define a temporal, directed, and weighted network, where nodes represent countries and edges represent votes.

Country participation varies over time. To ensure consistency, we restrict the analysis to the period 1995--2015 and consider only votes cast during the final stage.

\subsubsection{Preference Normalization}
To isolate voting preferences beyond overall performance, following~\cite{Dekker2007}, we normalize votes as follows. For each year $t$, let 
\[ 
    w_{ab}^{(t)} \quad \text{ and } \quad  \bar{w}_{b}^{(t)}
\]
respectively denote the number of points assigned by country $a$ to country $b$, and the average number of points received by country $b$ over all voting countries.
We define the preference weight:
\[ 
    \tilde{w}_{ab}^{(t)} = w_{ab}^{(t)} - \bar{w}_{b}^{(t)}.
\]
Only positive values are retained, i.e.,
\[ 
    \hat{w}_{ab}^{(t)} = \max\big(0, \tilde{w}_{ab}^{(t)}\big).
\]
This transformation aims to remove the baseline effect of song performance and highlight relative voting affinities between countries. 
The resulting network is temporal, directed, and weighted, with edges encoding positive preference weights.

\subsubsection{Network Variants}
To assess the impact of modeling choices, we construct four network representations: (a) Directed / weighted: original normalized network; (b) Directed / unweighted: all nonzero weights set to 1; (c) Undirected / weighted: reciprocal edges are merged into a single undirected edge with weight; (d) Undirected / unweighted: same as above, with all weights set to 1.

\subsubsection{LAGO Configuration}
LAGO is applied using the Mean Membership expectation, with parameters $\gamma = 4$ and $\omega = 4$. Since the defaults $\gamma = \omega = 1$ returned many small communities, we increase them in order to uncover structures relevant for visual analysis. STEM refinement is performed after each iteration of the core loop, and the fast exploration mode is enabled. For each configuration, the algorithm is run 10 times, and the partition with the highest L-Modularity is retained.

\subsubsection{Results}
\begin{figure}[htbp]
    \centering
    \begin{minipage}[b]{0.24\linewidth}
        \centering
        \includegraphics[width=\linewidth]{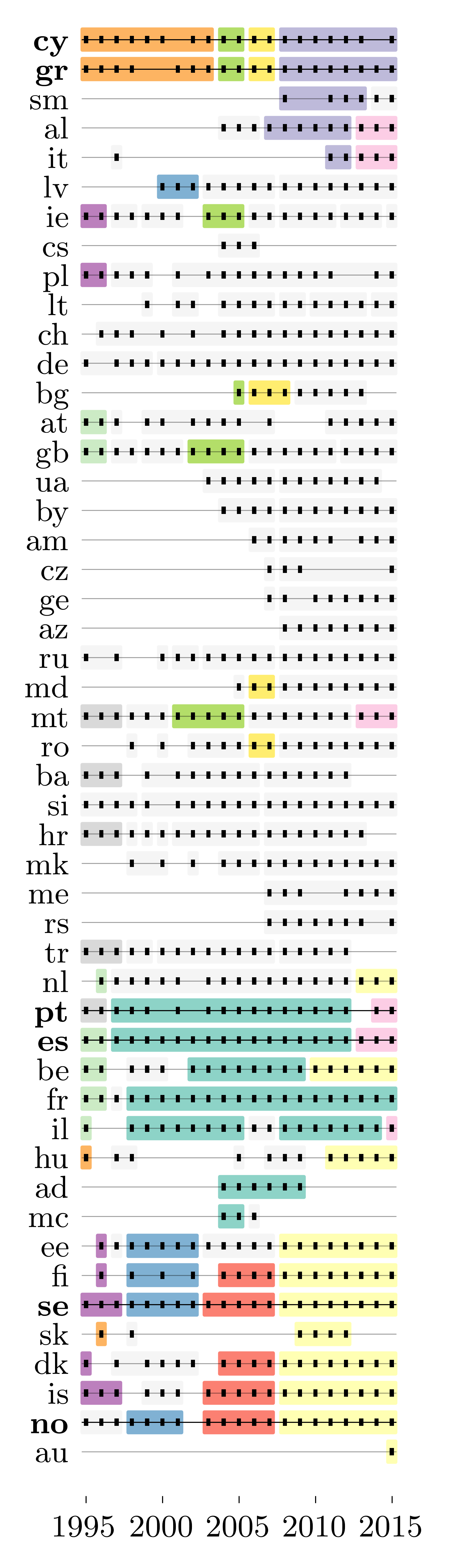}
        \vspace{2pt}
        \small (a) Dir./wt.
    \end{minipage}
    \hfill
    \begin{minipage}[b]{0.24\linewidth}
        \centering
        \includegraphics[width=\linewidth]{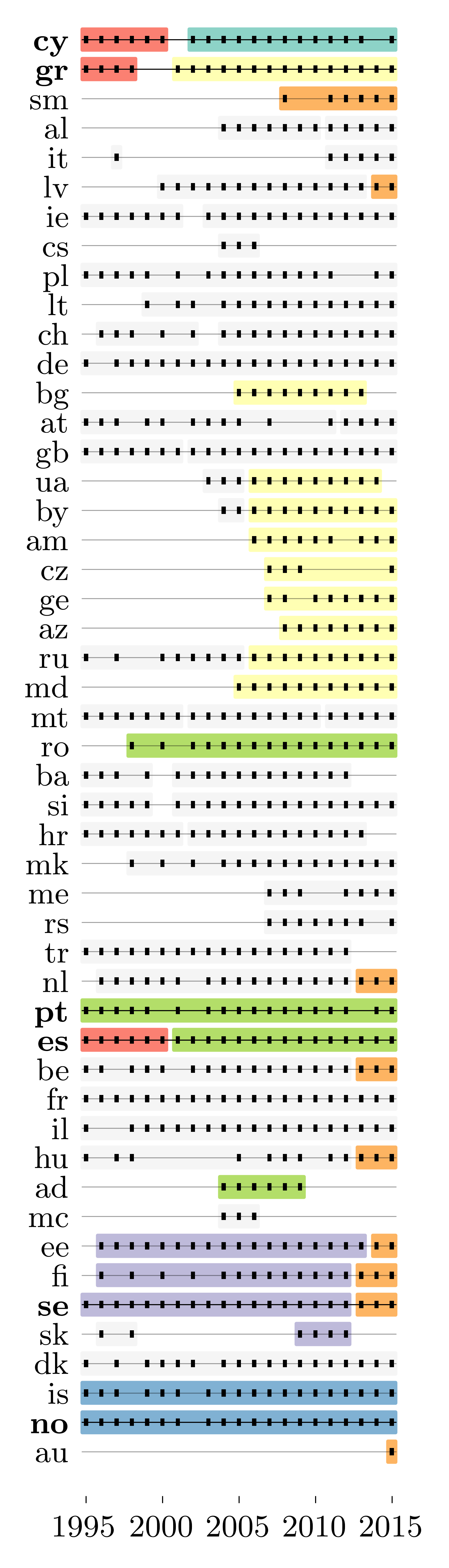}
        \vspace{2pt}
        \small (b) Dir./unwt.
    \end{minipage}
    \hfill
    \begin{minipage}[b]{0.24\linewidth}
        \centering
        \includegraphics[width=\linewidth]{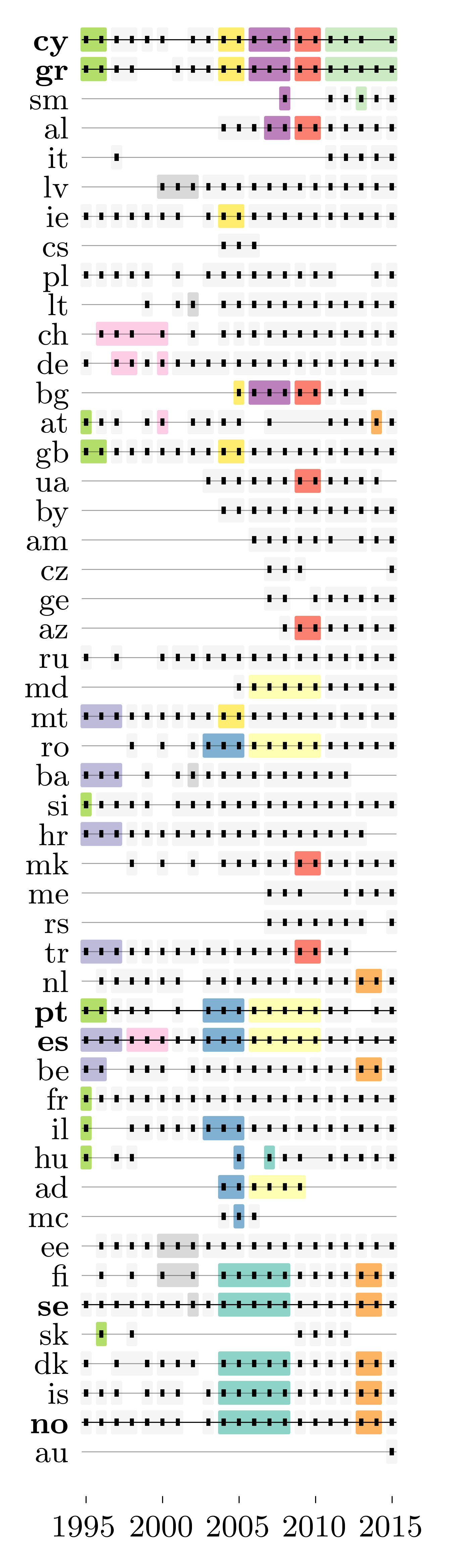}
        \vspace{2pt}
        \small (c) Undir./wt.
    \end{minipage}
    \hfill
    \begin{minipage}[b]{0.24\linewidth}
        \centering
        \includegraphics[width=\linewidth]{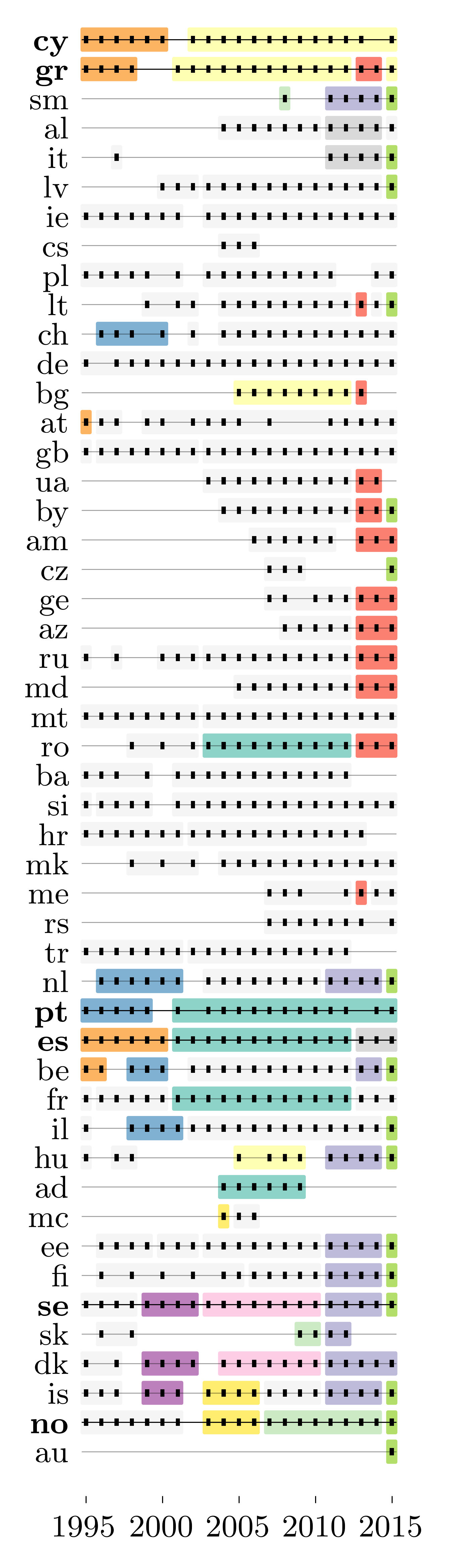}
        \vspace{2pt}
        \small (d) Undir./unwt.
    \end{minipage}

    \vspace{4pt} 
    \caption{\textbf{Temporal Eurovision voting network under four modeling configurations.} 
    Countries are shown on the y-axis (ISO 3166 codes) and years on the x-axis. Black squares indicate years in which a country participated in the vote. Edges are omitted for clarity. Colors represent dynamic communities detected by LAGO; only the 12 most present communities, corresponding to bold node labels, are shown in color, while all others are displayed in light grey. Comparing the four configurations highlights the impact of modeling choices (directed/undirected and weighted/unweighted) on the resulting community structure.}
    \label{fig:eurovision}
\end{figure}

The results are shown in Fig.~\ref{fig:eurovision}. Countries are denoted using ISO 3166-1 alpha-2 codes. The bold codes in the figure correspond to countries commented in the following; for example, \textbf{es} (Spain), \textbf{pt} (Portugal), \textbf{cy} (Cyprus), \textbf{gr} (Greece), \textbf{se} (Sweden), and \textbf{no} (Norway).

We observe substantial differences across the four configurations. These differences are not attributable to optimization variability, as multiple runs consistently converge to similar high-scoring solutions.

Several detected communities align with known voting patterns reported in prior work. In particular~\cite{melodyeurope}, the pairs (Spain, Portugal), (Cyprus, Greece), and (Sweden, Norway) are recurrently grouped together in the directed and weighted configuration. 

In contrast, the unweighted representations fail to consistently recover some of these associations, notably the Sweden--Norway pair. This suggests that edge weights carry important information about the strength of voting preferences.

Finally, undirected representations tend to produce communities with shorter temporal persistence. This is consistent with the change in the L-Modularity formulation when directionality is removed: the degree product $k_u k_v$ replaces $k_u^{\mathrm{out}} k_v^{\mathrm{in}}$, which for asymmetric voting profiles reshapes the null expectation and makes sustained co-membership harder to justify statistically.

\subsection{A Transportation System: Citi Bike}\label{sec:citibike}

Transportation systems are commonly modeled as networks, where nodes represent locations and edges capture flows of people or goods \cite{survey_chen_2025,spatiotemporal_rakhmangulov_2025,complexnetworkbased_zhang_2022}, among others. Applying community detection methods to such data reveals underlying mobility structures and their potential evolution over time \cite{dynamic_zhao_2023,revealing_xie_2021,identification_yildirimoglu_2018}. Link streams with delayed interactions are particularly well suited for modeling such temporal data.
The goal here is not to provide a detailed analysis of transportation patterns, but rather to illustrate the relevance of our approach on real-world temporal network data.

In this section, we apply LAGO to bike trips between stations of the Citi Bike system, New York City's public bicycle-sharing network. The dataset\footnote{\url{https://s3.amazonaws.com/tripdata/index.html}} provides station identifiers together with departure and arrival timestamps for each trip.

\subsubsection{Modeling Choices}

Trips can be modeled in two distinct ways. The most direct representation uses delayed interactions: each trip is an interaction starting at the departure time and ending at the arrival time, linking the origin and destination stations. Alternatively, trips can be modeled as continuous interactions, where the two stations are considered connected throughout the trip duration. This second representation assumes that stations are temporarily coupled while a bike is in transit, capturing ongoing system-level connectivity.
In both cases, nodes correspond to stations and interactions correspond to trips. The objective is to show that generalized L-Modularity optimization is able to discover meaningful communities in these two formalisms directly with LAGO, and to show that the modeling choices indeed have an effect on the results.

The delayed link stream is constructed directly from the data, with each trip represented by a single delayed interaction. The continuous link stream is obtained by assigning to each trip a continuous interaction spanning its duration. In this case, each interaction is weighted by the inverse of its duration, ensuring that node degrees remain identical across the delayed and continuous representations. Consequently, for a fixed community structure, longitudinal expectations and the temporal penalty term are preserved. The only difference lies in how intra-community interaction weights are accumulated. This normalization ensures that L-Modularity values remain comparable across the two models, allowing for a meaningful comparison of the resulting dynamic communities.

We investigate how the choice between delayed and continuous interaction models affects the structure and stability of dynamic communities.

\subsubsection{Data Filtering}

We restrict the analysis to the time window from 6:30 to 11:30 on 2025-11-10 (Monday), corresponding to morning commuting hours when usage is structured and intense. Only trips lasting at least 15 minutes are retained. Stations with the lowest degrees are iteratively removed until fewer than 50 stations remain.

These filtering steps reduce noise, remove marginal nodes, and limit the size of the temporal network to enable visual inspection, while emphasizing delayed interactions by focusing on relatively long trips.

\subsubsection{LAGO Configuration}

LAGO is applied using the Mean Membership expectation.
The parameters $\gamma$ and $\omega$ are both set to the default value of $1$. STEM refinement is performed after each iteration of the core loop, and the fast exploration mode is enabled. Each experiment is repeated 10 times, and the solution of highest L-Modularity is retained.

\subsubsection{Results}
\begin{figure}[htbp]
    \centering
    \begin{minipage}[b]{\linewidth}
        \centering
        \includegraphics[width=\linewidth]{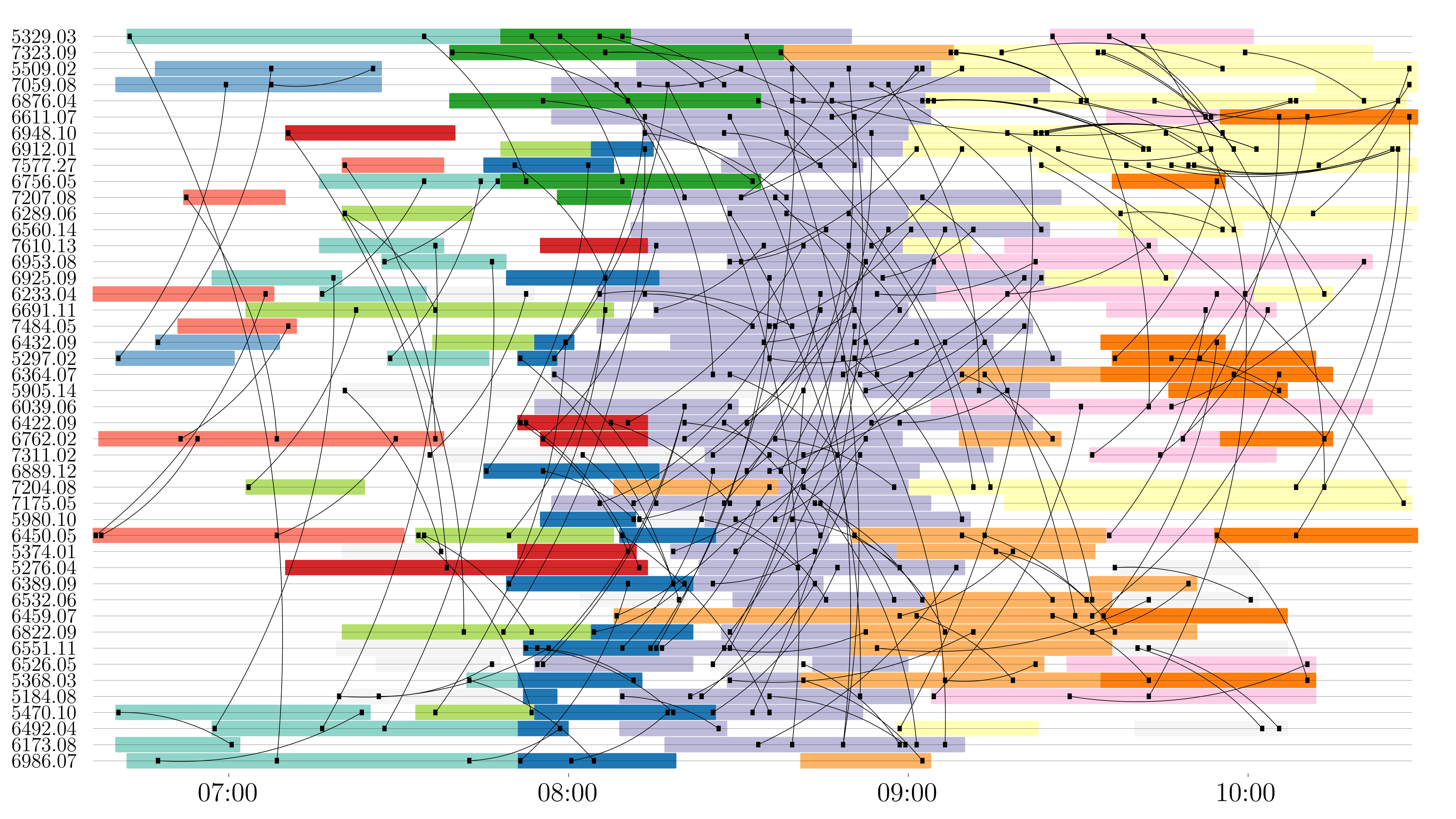}
        \vspace{2pt}
        \small (a) 15 communities detected using continuous interactions.
    \end{minipage}

    \vspace{3mm} 

    \begin{minipage}[b]{\linewidth}
        \centering
        \includegraphics[width=\linewidth]{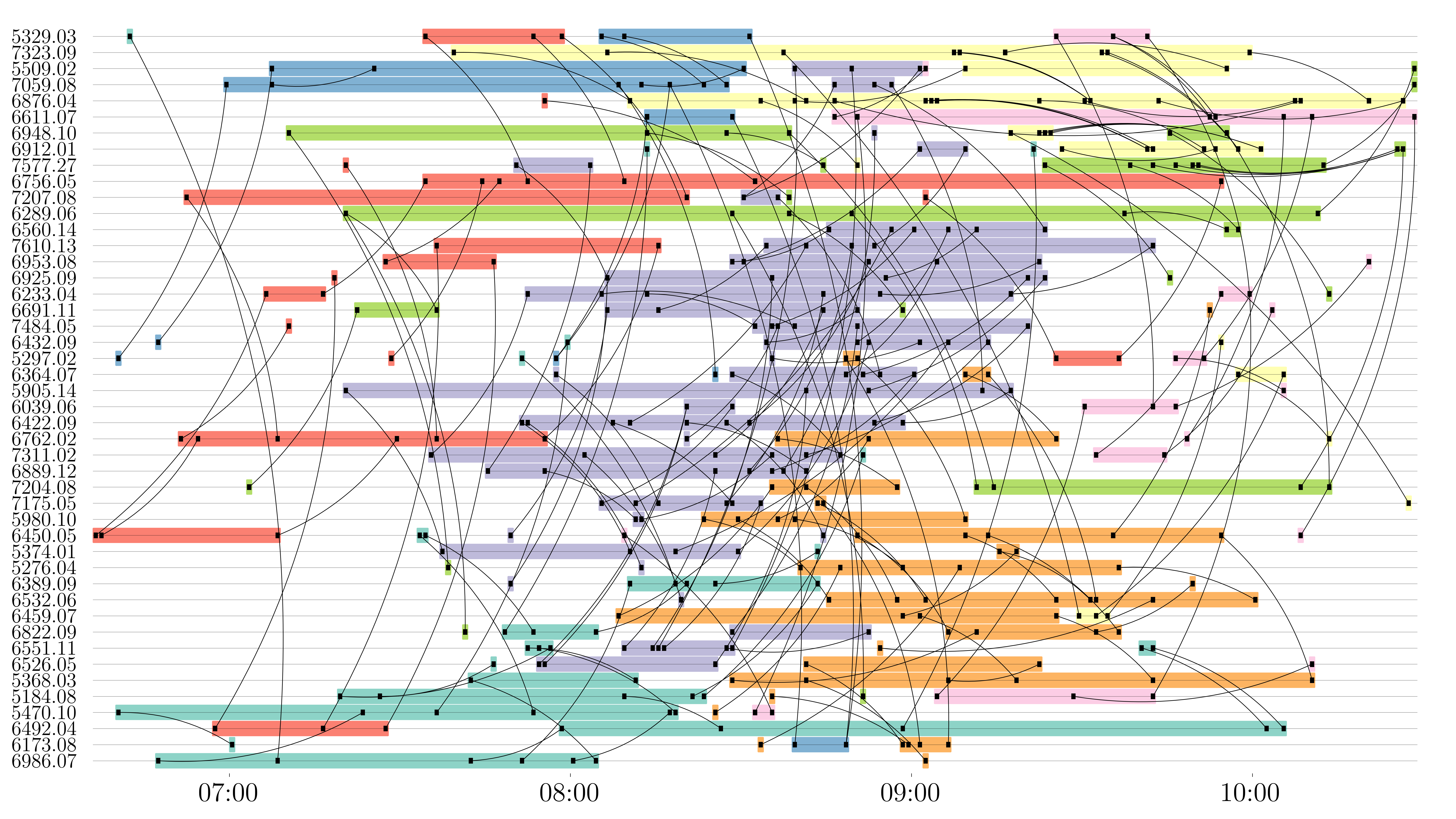}
        \vspace{2pt}
        \small (b) 7 communities detected using delayed interactions.
    \end{minipage}

    \vspace{4pt} 
    \caption{\textbf{Citi Bike trips on the morning of 2025-11-10 under two link stream models.} 
    Horizontal lines correspond to the nodes, i.e., docking stations. Colors indicate dynamic communities identified by LAGO. For clarity, only the 12 most present communities are colored; others are shown in light grey. Edges (trips) are displayed as delayed interactions for readability. Results differ between the two representations.}
    \label{fig:citibike}
\end{figure}

Figure~\ref{fig:citibike} presents the communities obtained under both models. The number of detected communities differs significantly (15 vs 7), highlighting the impact of the interaction model.

Continuous and delayed interaction models produce distinct temporal patterns in dynamic community structure. Continuous interactions reduce isolated time-node memberships and distribute the influence of interactions over their duration, resulting in smoother transitions between communities and more temporally coherent node assignments.

Delayed interactions emphasize temporally localized events, producing communities that can persist longer and allowing nodes to interleave across multiple dynamic communities. Under this model, some communities span nearly the entire lifespan of the network and include nodes that are not present simultaneously (e.g., yellow, red, and turquoise communities), while other nodes belong only briefly (e.g., node $6173.08$, second from the bottom, which switches from turquoise to orange to blue and back to orange). Such patterns, enabled by delayed interactions, provide novel insight into fine-grained temporal variations in transportation behavior.

Overall, continuous interactions yield smoother, more coherent community structures, whereas delayed interactions result in finer, more fragmented communities that capture localized temporal dynamics. By systematically considering delayed interactions, it is possible to characterize temporal patterns that are not observable in conventional continuous models.

\subsection{A Bipartite Network: Bluesky Messages}\label{sec:bluesky}

Online social networks can be modeled as temporal multipartite networks, where nodes represent heterogeneous entities (e.g., users and hashtags), and edges correspond to timestamped interactions. Community detection methods applied to such data reveal latent groups and the temporal evolution of topics or user interests~\cite{characterizing_cruickshank_2020,community_azaouzi_2019,community_bedi_2016}. However, many approaches rely on simplifying assumptions such as time aggregation or network projection, potentially obscuring temporal or structural information.

In this section, we model Bluesky messages as a bipartite link stream. Nodes correspond to users and hashtags, and interactions encode the publication of a hashtag by a user at a given time. We then apply LAGO to uncover dynamic communities. Our objective is illustrative: we aim to demonstrate the capacity of the method to directly handle this type of graph, rather than to provide an exhaustive social network analysis.

Bluesky is an online social network where users publish short messages that may include hashtags. Hashtags are user-defined keywords prefixed by the symbol \#. They act as explicit markers of content and facilitate topic-based organization. Because they directly reflect the intended topic of a message, they are well-suited for studying topic dynamics and community formation.

\subsubsection{Modeling Choices}

The dataset, introduced by Failla and Rossetti~\cite{blueskytonight}, was collected between February and May 2024 and comprises in total approximately $4$ million users and $235$ million posts. We use a subset of this data, provided by the authors, focusing on some topics (Ukraine war and Science).

\begin{figure}[!htb]
    \centering
    \includegraphics[width=\textwidth]{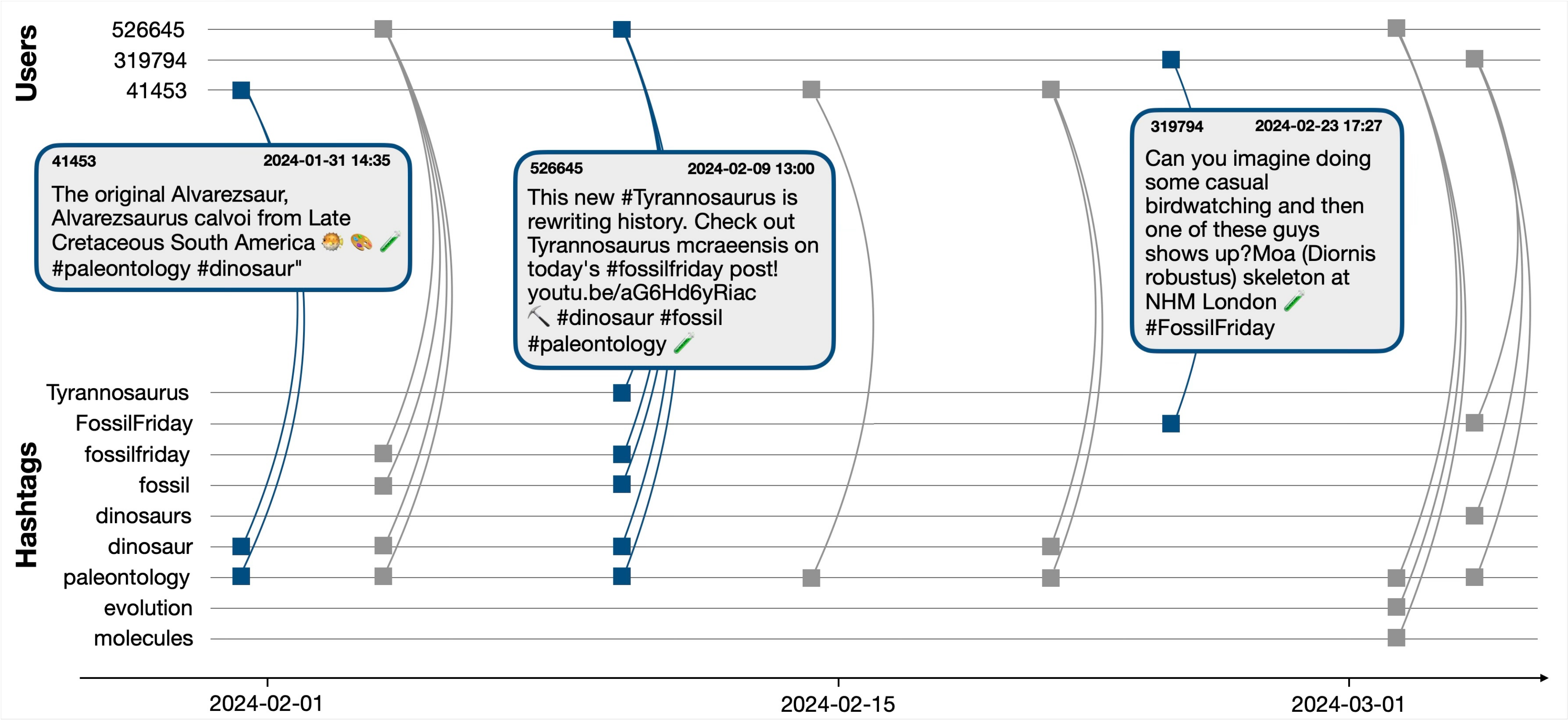}
    \caption{\textbf{Bluesky data representation.} Three users publish eight posts over time, involving nine distinct hashtags. Each post generates one or more interactions between the posting user and the hashtags it contains, all timestamped at the time of publication.}
    \label{fig:bluesky:model}
\end{figure}

We model the data as a bipartite temporal network by defining interactions of the form $(u, v, t)$, where a user $u$ publishes a post containing the hashtag $v$ at time $t$. By construction, interactions occur exclusively between users and hashtags, resulting in a bipartite structure, as illustrated in Fig.~\ref{fig:bluesky:model}.

Applying LAGO to this network enables the identification of (i) groups of hashtags co-used over time, revealing topic dynamics, and (ii) groups of users sharing coherent hashtag usage, capturing evolving user communities.

\subsubsection{Data Filtering}
We remove posts containing more than 10 hashtags, as they are likely to correspond to noise or spam. We then iteratively filter out users and hashtags with the lowest degrees. These filtering steps reduce noise, remove marginal nodes, and limit the size of the temporal network to improve interpretability.

\subsubsection{LAGO Configuration}
LAGO is applied using the Joint Membership expectation. Parameters are set to the default values $\gamma = 1$ and $\omega = 1$. STEM refinement is performed after each iteration of the core loop, and the fast exploration mode is enabled. The algorithm is run 10 times, and the solution with the highest L-Modularity is retained.

\subsubsection{Results}
\begin{figure}[htbp]
    \centering
    \begin{minipage}[b]{\linewidth}
        \centering
        \includegraphics[width=\linewidth]{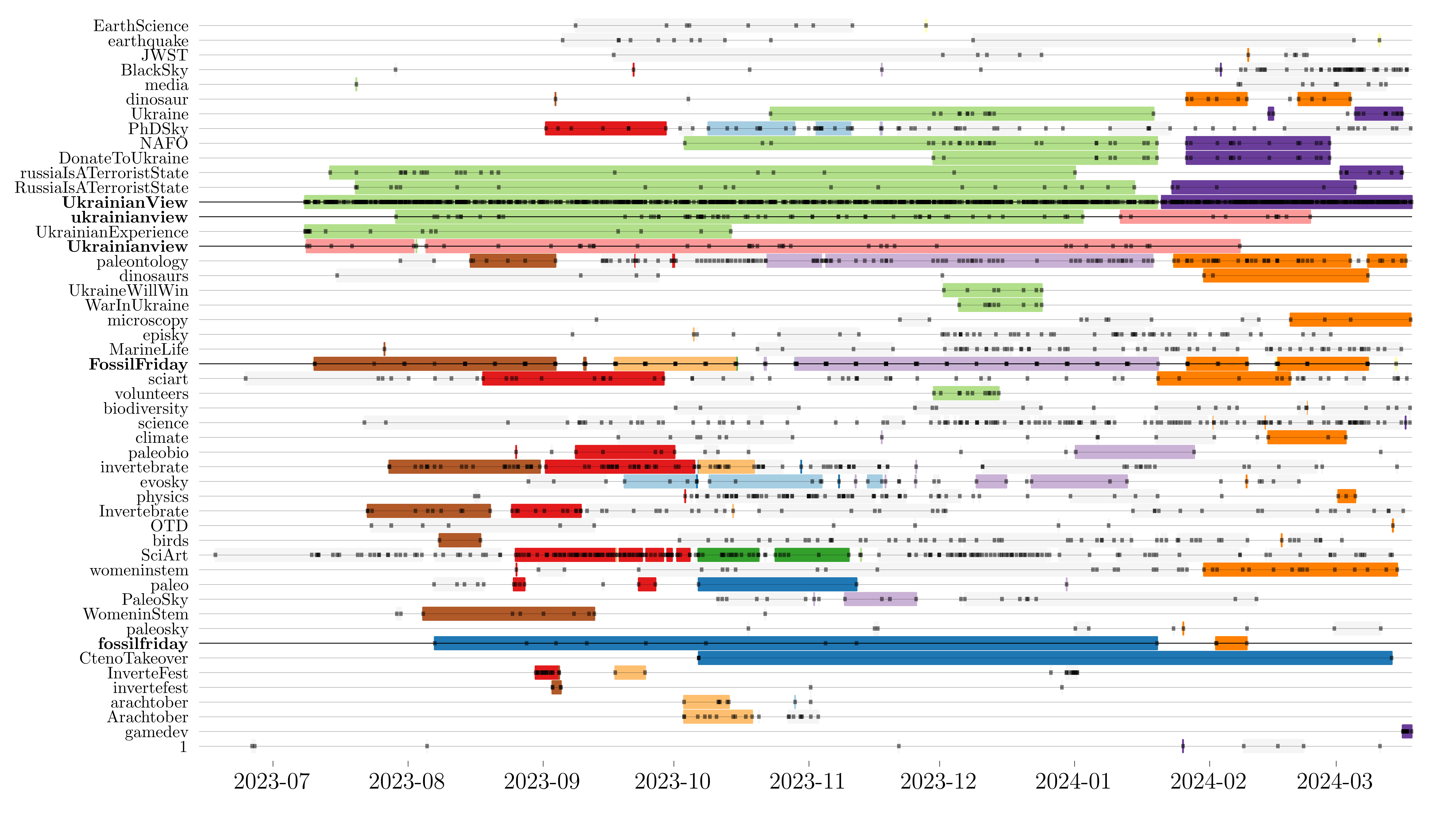}
        \vspace{2pt}
        \small (a) Temporal usage of 50 hashtags.
    \end{minipage}

    \vspace{3mm}

    \begin{minipage}[b]{\linewidth}
        \centering
        \includegraphics[width=\linewidth]{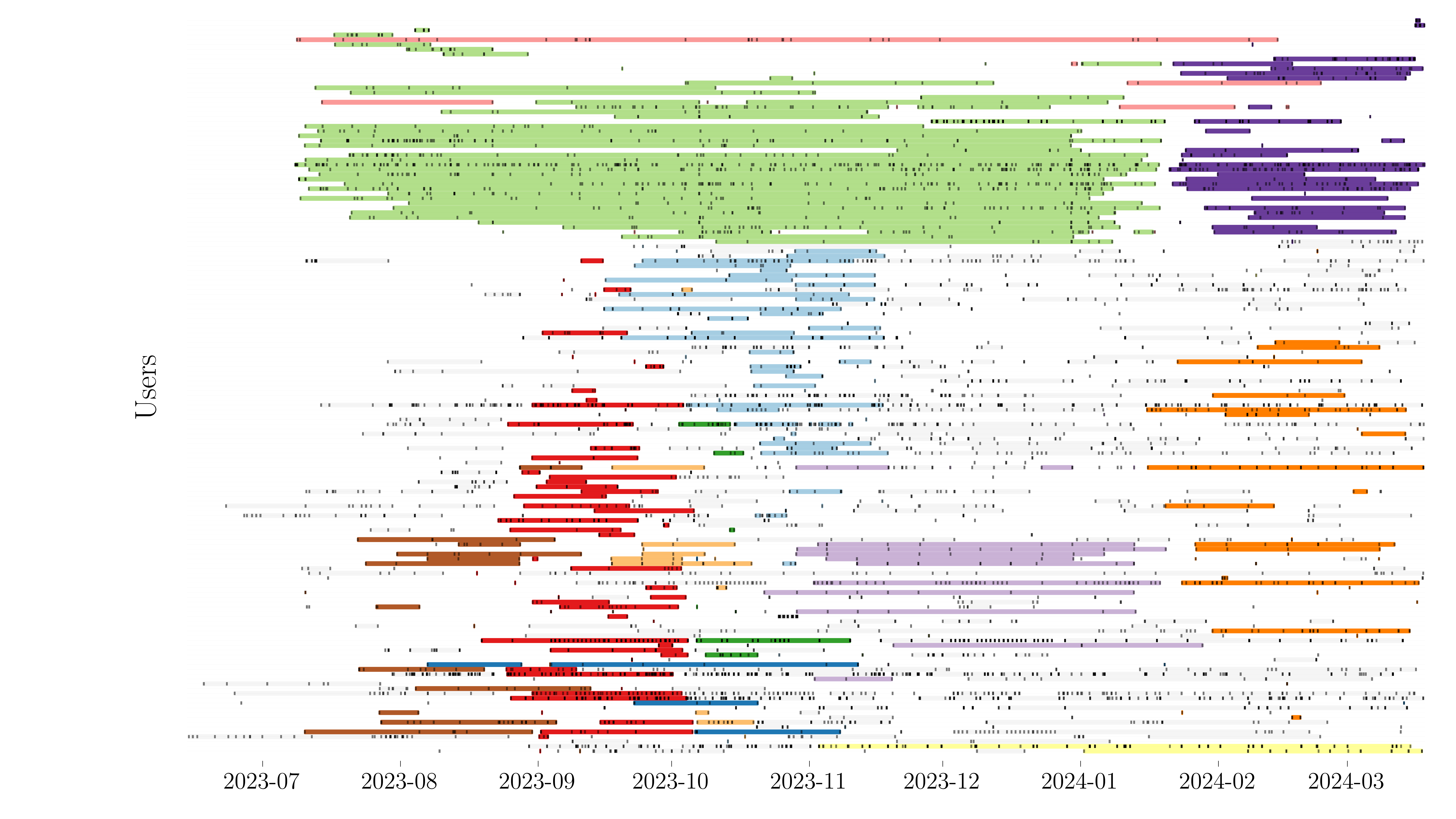}
        \vspace{2pt}
        \small (b) Temporal activity of 158 users.
    \end{minipage}

    \vspace{4pt} --
    \caption{\textbf{Visualization of LAGO results on the Bluesky bipartite temporal network.} 
    Black squares indicate activity—hashtag usage or user publication—at a given time. Colors represent dynamic communities detected by LAGO. Only communities associated with the bolded hashtags are colored, while all others are displayed in light gray. For readability, only subsets of users and hashtags most related to the highlighted communities are shown.}
    \label{fig:bluesky}
\end{figure}

Figure~\ref{fig:bluesky}.a displays communities of hashtags over time. Each point corresponds to an occurrence of a hashtag in a message. We focus on two sets of semantically equivalent variants: (i) \textit{UkrainianView}, \textit{ukrainianview}, and \textit{Ukrainianview}; and (ii) \textit{FossilFriday} and \textit{fossilfriday}.

One might expect such variants to belong to the same community. However, this is not systematically observed, suggesting heterogeneous usage across user groups.

For group (i), the variants are distributed across three communities (light green, purple, and pink). The light green community appears early and dominates. The purple community emerges in the final quarter of the observation period, possibly as a restructuring of the former. The pink community is smaller and more isolated. Around early February 2024, a transition occurs: \textit{ukrainianview} shifts to the pink community, while the remaining variants move from the light green to the purple community.

This transition is better understood from the user perspective: approximately one-third of the users in the light green community stop using these hashtags, while few new users adopt them.

For group (ii), multiple user communities engage with the same topic over time. Most exhibit high turnover (e.g., brown, light orange, purple, and orange communities). In contrast, the blue community remains stable and predominantly uses \textit{FossilFriday}, while others favor the lowercase form \textit{fossilfriday}. The associated hashtags evolve over time, including \textit{WomeninStem}, \textit{Arachtober}, \textit{paleontology}, and \textit{climate}, whereas others such as \textit{invertebrate} remain consistently present.

\section{Discussion}\label{sec:discussion}

We introduced a unified framework for dynamic community detection in temporal networks that generalizes both Longitudinal Modularity and the LAGO algorithm to weighted, directed, and multipartite interactions, as well as instantaneous, continuous, and delayed temporal modalities.

The experiments on three real-world datasets demonstrate that LAGO optimizes the generalized L-Modularity across diverse interaction types, producing communities that are interpretable in light of domain knowledge. Crucially, the results show that \emph{modeling choices matter}: retaining edge weights and directions in the Eurovision network better recovers known voting blocs; choosing between delayed and continuous models on Citi Bike produces markedly different community structures; and the bipartite formulation on Bluesky captures joint user--hashtag dynamics that a unipartite projection would lose. These findings support the central claim that avoiding destructive transformations preserves information relevant to community structure.

The current contribution still has limitations that could be explored in future works: parameters $\gamma$ and $\omega$ jointly control community granularity and temporal stability, but a principled selection strategy is still lacking. For interval-based interactions, the segmentation step can drastically increase the number of atomic elements. 
Finally, the temporal penalty treats all community switches equally regardless of interaction weight.

Several directions extend naturally from this work: investigating how the resolution limit interacts with the temporal dimension to guide a priori tuning; designing temporally heterogeneous null models (e.g., accounting for circadian or bursty patterns) better suited to some real-world datasets; adapting LAGO to streaming settings for incremental community updates; and extending the framework to additional network features such as signed interactions, higher-order structures, or overlapping communities.

In summary, the generalized L-Modularity provides a single quality function encompassing all considered interaction features, and LAGO optimizes it directly on raw temporal data. By removing the need for destructive transformations, the framework enables richer and more faithful analyses of complex temporal systems.

\backmatter

\bmhead{Supplementary information}

All code necessary to reproduce the experiments is available at:
\url{https://doi.org/10.5281/zenodo.19556274}.
An open-source Python library to apply LAGO and compute L-Modularity is available at:
\url{https://github.com/fondationsahar/dynamic\_community\_detection}.

\bmhead{Acknowledgements}

The authors thank SAHAR for financing this project.

\section*{Declarations}

\bmhead{Funding}
This work was supported by SAHAR.

\bmhead{Conflict of interest}
The authors declare that they have no conflict of interest.

\bmhead{Ethics approval}
Not applicable.

\bmhead{Consent to participate}
Not applicable.

\bmhead{Consent for publication}
Not applicable.

\bmhead{Data availability}
The datasets generated and analyzed during the current study are available at:
\url{https://doi.org/10.5281/zenodo.19556274}.

\bmhead{Materials availability}
Not applicable.

\bmhead{Code availability}
All code necessary to reproduce the experiments is available at:
\url{https://doi.org/10.5281/zenodo.19556274}.
An open-source Python library to apply LAGO and compute L-Modularity is available at:
\url{https://github.com/fondationsahar/dynamic\_community\_detection}.

\bmhead{Author contributions}
AB supervised the research. VB and RC conceived and designed the study. VB performed the experiments. VB and RC wrote the manuscript. All authors read and approved the final manuscript.

\bmhead{Declaration on the use of AI tools}
Artificial intelligence (AI) tools were used for language refinement, grammar correction, text formatting, and manuscript editing assistance. All outputs generated by these tools were reviewed, verified, and edited by the authors. The authors take full responsibility for the content of the final manuscript.

\bibliography{refs}

\end{document}